\documentclass[prd,twocolumn,showpacs,superscriptaddress]{revtex4}
\pdfoutput=1

\usepackage{amsfonts}
\usepackage{amsmath}
\usepackage{amssymb}
\usepackage{bm}
\usepackage{dcolumn}
\usepackage{epsfig}
\usepackage[T1]{fontenc}
\usepackage{graphicx}
\usepackage{graphics}
\usepackage{hyperref}
\usepackage[latin1]{inputenc}
\usepackage{latexsym}
\usepackage{multirow}
\usepackage{rotating}

\newcommand\be{\begin{equation}}
\newcommand\ba{\begin{eqnarray}}
\newcommand\ee{\end{equation}}
\newcommand\ea{\end{eqnarray}}

\newcommand{\mb}[1]{\mbox{\boldmath $#1$}}
\newcommand{\salto}[1]{\left[\,#1\,\right]^{}_{p}}
\newcommand{\nn}{\nonumber}
\newcommand{\met}{\mbox{g}}

\newcommand{\rsu}{r^{\ast}}

\newcommand{\singu}{{\mbox{\tiny S}}}
\newcommand{\regu}{{\mbox{\tiny R}}}
\newcommand{\Hor}{{\mbox{\tiny H}}}
\newcommand{\Inf}{{\mbox{\tiny I}}}

\newcommand{\MAX}{{\mbox{\tiny max}}}

\begin{document}

\title[An Efficient Pseudospectral Method for the Computation of the Self-force 
on a Charged Particle: Circular Geodesics around a Schwarzschild Black Hole]
{An Efficient Pseudospectral Method for the Computation of the Self-force 
on a Charged Particle: Circular Geodesics around a Schwarzschild Black Hole}

\author{Priscilla Ca\~nizares}
\affiliation{Institut de Ci\`encies de l'Espai (CSIC-IEEC), 
Facultat de Ci\`encies, Campus UAB, Torre C5 parells, 
E-08193 Bellaterra, Spain}

\author{Carlos F.~Sopuerta}
\affiliation{Institut de Ci\`encies de l'Espai (CSIC-IEEC), 
Facultat de Ci\`encies, Campus UAB, Torre C5 parells, 
E-08193 Bellaterra, Spain}

\date{\today}

\preprint{}

\begin{abstract}
The description of the inspiral of a stellar-mass compact object into a massive
black hole sitting at a galactic centre is a problem of major relevance for the
future space-based gravitational-wave observatory LISA (Laser Interferometer Space Antenna), 
as the signals from these systems will be buried in the data stream and accurate
gravitational-wave templates will be needed to extract them.
The main difficulty in describing these systems lies in the estimation of the
gravitational effects of the stellar-mass compact object on his own trajectory
around the massive black hole, which can be modeled as the action of a local
force, the {\em self-force}.
In this paper, we present a new time-domain numerical method for the computation of the 
self-force in a simplified model consisting of a charged scalar particle orbiting
a nonrotating black hole.  We use a multi-domain framework in such a way that the
particle is located at the interface between two domains so that the presence of the 
particle and its physical effects appear only through appropriate boundary conditions.
In this way we eliminate completely the presence of a small length scale associated
with the need of resolving the particle.  This technique also avoids the problems
associated with the impact of a low differentiability of the solution in the 
accuracy of the numerical computations.  The spatial discretization of the field
equations is done by using the pseudospectral collocation method and the time 
evolution, based on the method of lines, uses a Runge-Kutta solver.
We show how this special framework can provide very efficient and accurate computations 
in the time domain, which makes the technique amenable for the intensive computations
required in the astrophysically-relevant scenarios for LISA.
\end{abstract}


\maketitle

\section{Introduction}
\label{intro}
One of the main sources of gravitational radiation for the future ESA-NASA 
gravitational wave observatory, the Laser Interferometer Space Antenna (LISA)~\cite{LISA},
is the capture and inspiral of a stellar-mass compact object (SCO), with 
masses in the range $m = 1-50 M^{}_{\odot}$, into a massive black hole (MBH) in a
galactic centre, with masses in the range $M= 10^4-10^7 M_{\odot}$.  These systems are usually 
known as Extreme-Mass-Ratio Inspirals (EMRIs) since the mass ratios involved are in the range
$\mu=m/M \sim 10^{-7}-10^{-3}$.  During the inspiral phase the system is driven by 
the emission of gravitational radiation, and hence there is loss of energy and angular 
momentum that makes the orbit shrink until the SCO plunges into the MBH.  
It is expected that LISA will be able to detect 
$10-10^3$ EMRI$/yr$~\cite{Gair:2004iv,Hopman:2006xn} up to distances 
within $1\,Gpc$~\cite{Sigurdsson:1996uz} (see also~\cite{AmaroSeoane:2007aw}
for more details on the astrophysics of EMRIs). 
Recently, it has been suggested~\cite{Brown06,Mandel:2007hi} that inspirals of SCOs
into intermediate-mass black holes, with masses in the range $50 - 350\, M_\odot$ and
presumably located in globular clusters, could be detected by
future second-generation ground interferometers like Advanced LIGO~\cite{AdvLIGO}
and Advanced VIRGO~\cite{AdvVIRGO}.  The mass ratios are in the range $10^{-3} - 10^{-1}$,
and in consequence they are called Intermediate-Mass-Ratio Inspirals (IMRIs).  It is 
expected that techniques to describe EMRIs may also been used for IMRIs, at least to a 
certain degree of precision.

During the long inspiral, an EMRI will spend many cycles inside the LISA band, 
of the order of $\sim 10^5$ during the last year before plunge into the MBH~\cite{Finn:2000sy}.
However, the gravitational-wave signals from EMRIs will be buried in the LISA data stream
with the instrumental noise and the gravitational wave foreground (produced by compact 
binaries in the LISA band).  To extract these signals and the relevant physical parameters 
that characterize them, we need to have a very precise {\em a priori} theoretical knowledge 
of the gravitational waveforms.  This means to describe the inspiral of the SCO taking
into account the gravitational backreaction, that is, the influence of the SCO gravitational
field on its own motion.  This is an interesting but difficult theoretical problem, and
different methods have been developed to solve 
it (see~\cite{Poisson:2004lr,Tanaka:2005ue,Glampedakis:2005hs}).

While techniques for constructing templates good enough for detection are getting ready, 
mainly based on the use of the adiabatic
approximation~\cite{Hughes:2005qb,Drasco:2005kz,Sago:2005gd,Ganz:2007rf},
methods to build templates good enough for extraction of physical information are not yet fully 
developed. The main difficulty being that one requires a more precise treatment of the self-gravity 
of the SCO and its impact on the gravitational waveform. In relation to this fact, there is 
currently a significant activity on the study of the accuracy of the different types of adiabatic
approximations that have been
introduced~\cite{Mino:2003yg,Pound:2005fs,Pound:2007ti,Mino:2007ft,Hinderer:2008dm}.
On the other hand, the {\em self-force} approach to the equations of 
motion~\cite{Mino:1997nk,Quinn:1997am} (see also~\cite{Detweiler:2002mi,Poisson:2004lr,Gralla:2008fg})
is an step forward to a precise estimation of the radiation-reaction effects and in
consequence, towards the construction of accurate waveform templates.
In this approach the backreaction effects on the SCO are described as the action
of a local force, the {\em self-force}, which can be computed in terms of the 
perturbations generated by the SCO with respect the MBH background spacetime.  
In practice, to compute the self-force one needs to regularize the perturbations,
similarly as it happens in electromagnetism~\cite{Barut:1980ao,Jackson:99jd}.
To that end, a {\em mode sum} regularization scheme has been designed (in the gravitational case 
it has been formulated in the Lorenz
gauge)~\cite{Barack:1999wf,Barack:2000eh,Barack:2001bw,Mino:2001mq,Barack:2001gx,Barack:2002mha,Detweiler:2002gi,Haas:2006ne}.
It tells us how to subtract, mode by mode (for a Schwarzschild MBH the modes correspond to a harmonic 
decomposition of the perturbations),
the singular part of the perturbations that does not contribute to the self-force.
Therefore, what we need is a method of computing the full retarded solution of the
perturbative equations for applying the mode-sum regularization scheme and obtain
in this way the self-force.  
The perturbative field equations are a set of ten linear partial differential 
equations (PDEs) for the metric perturbations $h^{}_{\mu\nu}$ ($h^{}_{\mu\nu} = 
\met^{}_{\mu\nu} - \bar{\met}^{}_{\mu\nu}$ at linear order, where $\met^{}_{\mu\nu}$
is the spacetime metric and $\bar{\met}^{}_{\mu\nu}$ is the background metric 
describing the MBH), which only in certain gauges (e.g., the Regge-Wheeler gauge~\cite{Regge:1957rw})
can be decoupled.  In any case, to solve them completely we need to resort to 
numerical methods.  Following the initial studies of black hole quasinormal modes,
frequency methods were used 
successfully ~\cite{Davis:1972pa,Detweiler:1978aj,Detweiler:1979ds,Cutler:1994pb,Poisson:1995vs,Poisson:1997ad},
and it was found they provide accurate results for EMRIs with moderate eccentricities.
However, the frequency domain approach has more difficulties with highly eccentric orbits, which are 
of interest for LISA, since one has to sum over a large number of modes to obtain a good accuracy, 
and convergence may be an issue. This has opened the door to time-domain methods, which are not 
affected much by the eccentricity of the orbit and may be more efficient for the case of 
high-eccentricity EMRIs.  In the last years there has been an intense activity on this 
front, both for a nonrotating 
background~\cite{Martel:2003jj,Barack:2005nr,Sopuerta:2005gz,Haas:2006ne,Haas:2007kz,Vega:2007mc,Barack:2007tm}, 
and for a rotating 
background~\cite{Burko:2006ua,Sundararajan:2007jg,Sundararajan:2008zm}.  
The main drawbacks of time-domain methods have mainly two origins: (i) The fact that one has to
resolve very different physical scales (both spatial and temporal) present in the problem due to the
extreme mass ratios involved  (see, e.g.~\cite{Sopuerta:2005rd}).  That is, using a standard numerical 
discretization of the problem we are led to resolve the typical gravitational wavelengths 
(comparable to the size of the MBH) and, at the same time, scales in the vicinity of the SCO, 
which are crucial for evaluating the self-force.  This translates in a demanding requirement 
of computational resources.  (ii) The fact that the
SCO is described as a point-like object. This introduces Dirac delta distributions in the SCO
energy-momentum distribution that lead to loss of differentiability in the solution of
the perturbative field equations.  This fact can degrade the convergence properties of the numerical
algorithms used.  Moreover, such a localized distribution of matter can also introduce spurious 
high-frequency modes that contaminate the numerical solution and, in consequence, degrade its accuracy.

Recently, there have been different proposals to improve the performance of time domain
methods.  Barack and Goldbourn~\cite{Barack:2007jh} have introduced a new technique to 
compute the scalar field generated by a pointlike scalar charge orbiting a black hole. This technique 
consists in subtracting from each azimuthal mode (in the Kerr geometry the field equations are not 
fully separable in the time domain and one has to tackle them in 2+1 dimensions)
of the retarded field a piece that describes the singular behavior near the
particle. This is done through a careful analytical study of the scalar field near the particle,
using a {\em puncture} scheme which resembles the puncture model used for simulations in numerical
relativity~\cite{Brandt:1997tf,Campanelli:2005dd,Baker:2005vv}.   This technique has been
extended to the electromagnetic and gravitational cases by Barack, Goldbourn and Sago~\cite{Barack:2007we}.
On the other hand, Vega and Detweiler~\cite{Vega:2007mc} have introduced another new method for regularizing
the solution of the field equations.  Their approach, tested on a simplified model of a charged
particle orbiting a nonrotating black hole, regularizes the retarded field itself by identifying
and removing first, in an analytical way, the singular part of the retarded field.  This alternative
approach to the mode-sum regularization scheme yields a finite and differentiable remainder from which 
the self-force can be computed. This remainder is the solution to a field equation with a nonsingular 
source, which avoid the problem (ii) above.  
Finally, Lousto and Nakano~\cite{Lousto:2008mb} have also introduced an analytical technique to
remove the particle singular behaviour.  Their method is global and also produces a well behaved
source for the field equations.

Whereas these new techniques help in dealing with problem (ii) above, they do not completely
solve the problem (i) since the regular source terms that these new schemes produce still have
associated with them a length scale (or, from the numerical point of view, there are still special 
spatial resolution requirements associated with those source terms).
In this paper, we introduce a new time-domain scheme towards the computation of the self-force
which, for the case of a nonrotating black hole, eliminates completely any length scale 
associated with the SCO.  This is done by using multiple subdomains and locating the particle 
in the interface between two of them (this has similarities with what was done in~\cite{Sopuerta:2005gz} 
using the finite element method, where in one of the numerical schemes proposed the particle 
was located between two elements).  In this way, 
the Dirac delta distributions do not appear in our equations and the presence of the SCO enters through 
the boundary conditions that communicate the solutions at the different subdomains.  As a consequence, we 
are solving wave-type equations with smooth solutions, which avoid the problems described in (ii) 
(preliminary results have been reported in~\cite{Canizares:2008dp}). 
Regarding (i), we just need to provide the numerical resolution to describe the field near 
the particle, but not the particle itself, which makes the computation much more efficient.  
Our numerical algorithms are based on the PseudoSpectral Collocation (PSC) method (see, e.g.~\cite{Boyd}),
which has been applied to numerical relativity~\cite{Grandclement:2007sb}, and recently it has
also been used in~\cite{Jung:2007zf} for one-dimensional head-on collisions of 
black-holes. And very recently, in~\cite{Field:2009kk}, a discontinuous Galerkin method 
has been introduced and some gravitational waveforms for extreme-mass-ratio binaries are
computed.  This work uses similar techniques to the ones introduced in~\cite{Sopuerta:2005gz}
and the ones that we present here.

In this paper, we describe a set of techniques and methods to use the PSC for the computation of 
the self-force on a charged scalar particle in circular orbits around a non-rotating black hole. 
We also show some results of the numerical implementation.  
The organization of the paper is as follows: In section~\ref{modeldescription} we introduce the basics
of the model of a charged scalar particle orbiting a nonrotating black hole, including the
basic formulae for the computation of the self-force via the mode-sum regularization scheme.
In section~\ref{timedomainframework} we introduce all the ingredients of a new time-domain 
numerical framework for the computation of the self-force in such scenario, from the mathematical
foundations to the practical implementation of the computations.
In section~\ref{results} we show the performance of a numerical code we have designed
to implement the new scheme, and results of the computation of the self-force, in particular
for the innermost stable circular orbit.  In section~\ref{discussion} we
draw conclusions from the results shown and discuss possible future
avenues in the development of these techniques for the simulations of
EMRIs in relevant physical situations.
Throughout this paper we use the metric signature $(-,+,+,+)$ and geometric
units in which $G = c = 1$.

\section{Description of the Model: Charged scalar particle orbiting a 
nonrotating MBH}  \label{modeldescription}

In this paper we present a new technique for the computation of the self-force.
In order to simplify things we focus in the particular case of a charged scalar
particle in circular geodesics around a non-rotating MBH.  In this simplified
model, the spacetime metric is not dynamical, in the sense that it is not affected
neither by the particle nor by the scalar field that it generates.  In our case,
since we are considering non-rotating BHs, the metric is the Schwarzschild metric,
which can be written as follows:
\begin{equation}
ds^2=f(-dt^2+d\rsu{}^{2})+r^2d\Omega^2\,,~
d\Omega^2=d\theta^2+\sin^2\theta d\varphi^2\,, \label{schmetric}
\end{equation}
where $(x^{\mu}) =(t,r,\theta,\varphi)$ are the so-called Schwarzschild coordinates,
$f(r)=1-{2M}/{r}$ (where $M$ is the BH mass), and $\rsu$ is the {\em tortoise} 
coordinate, given by:
\begin{equation}
\rsu = r + 2M \ln\left(\frac{r}{2M}-1\right)\,.
\end{equation}
In this geometry we assume there is a particle with scalar charge $q$, associated 
to a scalar field $\Phi(x^{\mu})$.  This particle is orbiting the BH, and in doing
so generates scalar field $\Phi$, which in turn influences the particle trajectory.
That is, the particle motion is affected by the field created by itself. In this way,
this model contains all the ingredients of the gravitational case, in which the 
particle motion is influenced by its own gravitational field.  

The equation for the scalar field is then (see, e.g.~\cite{Poisson:2004lr}):
\begin{equation}
\met^{\alpha\beta}\nabla^{}_{\alpha}\nabla^{}_{\beta}\Phi(x)=-4\pi \rho = 
-4\pi q\int^{}_{\gamma}d \tau\, 
\delta^{}_4(x,\textit{z}(\tau))  \,, \label{geo}
\end{equation}
that is, a wave-type equation with a source term that describes the particle 
energy density due to its scalar charge.  In this equation, the spacetime metric
$\met^{}_{\mu\nu}$ is the Schwarzschild metric~(\ref{schmetric}), $\nabla^{}_{\mu}$ denotes
the associated canonical connection; $\tau$ denotes proper time associated with
the particle along its timelike worldline $\gamma$, which we denote by $x^{\mu} = z^{\mu}(\tau)$; 
$\delta^{}_{4}(x,x')$ is the invariant Dirac functional in Schwarzschild spacetime, which is 
defined by the relation
\begin{equation}
\int^{}_{\gamma} d^{4}x \sqrt{-\met(x)} f(x)\delta^{}_{4}(x,x') = f(x')\,,
\end{equation}
and the equivalent one for the other argument.  In this relation, $\met$ denotes the
metric determinant.   Taking into account the properties of $\delta^{}_{4}(x,x')$, it
follows that the source term in the scalar field equation~(\ref{geo}) only has
support on the particle worldline $\gamma$.

The equations of motion for the particle trajectory ($x^{\mu} = z^{\mu}(\tau)$) that one would 
obtain from energy-momentum conservation are:
\begin{equation}
m\frac{du^{\mu}}{d\tau} = F^{\mu} = q (\met^{\mu\nu} + u^{\mu}u^{\nu})\nabla^{}_{\nu}\Phi\,, ~~
u^{\mu} = \frac{dz^{\mu}}{d\tau} \,, \label{particlemotion}
\end{equation}
where $m$ and $u^{\mu}$ are the particle mass and 4-velocity, respectively.  However, 
this expression is a formal one due to the fact that the force $F^{\mu}$ diverges on the particle 
worldline $\gamma$ (see~\cite{Quinn:2000wa} for a derivation of the {\em regularized} 
equations of motion).  An analysis of the solutions of (\ref{geo}) and (\ref{particlemotion})
reveals (see for details~\cite{Poisson:2004lr}) that the gradient of the field, 
$\nabla^{}_{\mu}\Phi$, can be split into two parts~\cite{Detweiler:2002mi}: A singular piece, 
$\Phi^{\singu}$, which contains the singular structure of the field and satisfies the same field 
equation, that is,
\begin{equation}
\met^{\alpha\beta}\nabla^{}_{\alpha}\nabla^{}_{\beta}\Phi^{\singu} =-4\pi \rho
 \,, \label{eqsingular}
\end{equation}
and a regular part, $\Phi^{\regu} = \Phi - \Phi^{\singu}$, which satisfies an homogeneous 
wave equation
\begin{equation}
\met^{\alpha\beta}\nabla^{}_{\alpha}\nabla^{}_{\beta}\Phi^{\regu} = 0\,,
\label{eqregular}
\end{equation}
and which is solely responsible of the deviation of the particle from geodesic motion
around the BH,
\begin{equation}
m\frac{du^{\mu}}{d\tau} =  q (\met^{\mu\nu} + u^{\mu}u^{\nu})\nabla^{}_{\nu}\Phi^{\regu}
\,. \label{regparticlemotion}
\end{equation}
This part is associated with the tail part of the scalar field.  This is the analogous 
equation to the {\em MiSaTaQuWa} equation of the gravitational case 
(see~\cite{Mino:1997nk,Quinn:1997am}), and $F^{\regu}_{\mu}=  
q (\met^{\mu\nu} + u^{\mu}u^{\nu})\nabla^{}_{\nu}\Phi^{\regu}$ is called the {\em
self-force} on the particle.

In order to solve the equation for the scalar field [Eq.~(\ref{geo})] it is very
convenient to take advantage of the spherical symmetry of the Schwarzschild spacetime 
and decompose $\Phi$ in scalar spherical harmonics, $Y^{m}_{\ell}(\theta,\varphi)$ 
(see Appendix~\ref{sphericalharmonics}), 
\begin{equation}
\Phi(x)=\sum_{\ell=0}^{\infty}\sum_{m=-\ell}^{\ell}\Phi^{m}_{\ell}(t,r)
Y^{m}_{\ell}(\theta,\varphi)\,,
\label{scalarfieldharmonics}
\end{equation}
where the harmonic numbers $(\ell,m)$ take the usual values:
$\ell=0,1,2,...,\infty$, and $m=-\ell,-\ell+1,...,\ell-1,\ell$.
Since we want to compute the self-force on the particle for geodesics
orbits, and since geodesic motion takes place on a plane in Schwarzschild geometry, 
we will assume, without loss of generality, that the plane is given by 
$\theta=\pi/2$.  We will also parameterize the motion of the particle 
in terms of the coordinate time $t$, instead of proper time $\tau$.
That is, the particle world-line, $\gamma$,  will be given by 
$(t,r^{}_{p}(t),\pi/2,\varphi^{}_{p}(t))\,$.  Taking this into account,
we can introduce the expansion~(\ref{scalarfieldharmonics}) into the scalar
field equation~(\ref{geo}) and find that the equations for the different
harmonic coefficients $\Phi^{m}_{\ell}(t,r)$ decouple and have the
form of a 1+1 wave-type equation:
\begin{eqnarray}
\left\lbrace  -\frac{\partial^2}{\partial t^2} + \frac{\partial^2}{\partial \rsu{}^{2}} 
-V^{}_{\ell}(r) \right\rbrace(r\Phi^{m}_{\ell})= S^{m}_{\ell}\delta (r-r^{}_{p}(t))\,, 
\label{master}
\end{eqnarray}
where $V^{}_{\ell}(r)$ is the Regge-Wheeler potential for scalar fields on the 
Schwarzschild geometry, given by
\begin{eqnarray}
V^{}_{\ell}(r) = f(r) \left[ \frac{\ell(\ell+1)}{r^2}+\frac{2M}{r^3}\right]\,,
\end{eqnarray}
and $S^{m}_{\ell}$ is the coefficient of the singular source term generated
by the particle:
\begin{eqnarray}
S^{m}_{\ell} = -\frac{4\pi qf^2(r^{}_{p})}{r^{}_{p}u^t}\,
\bar{Y}^{m}_{\ell}(\frac{\pi}{2},\varphi^{}_{p})\,,
\label{source}
\end{eqnarray}
where an overbar denotes complex conjugation.  

On a hypersurface $\{t=t_{o}\}$ we can prescribe initial data for $\Phi^m_{\ell}$, 
$({}_{o}\Phi^m_{\ell},{}_{o}\dot{\Phi}^m_{\ell}) = (\Phi^m_{\ell}(t_{o},r),
\partial_{t}\Phi^m_{\ell}(t_{o},r))$, and then find the corresponding solution
(given that $\Phi$ satisfies a wave equation, the problem is well-posed).
There are a couple of comments in order: First, the solution found in this way
corresponds to the $(\ell,m)$ contribution to the full retarded solution.  Second,
this solution will be finite and continuous at the particle location,
but it will not be differentiable at the particle location in the sense that
the radial derivative from the left and from the right of the particle yields
different values.  Moreover, the sum of the multipole coefficients over $\ell$
will diverge at the particle location.  This can be fixed by extracting, multipole
by multipole, the singular part of the scalar field.  Since we are interested
on regularizing the self-force, which is defined at the particle location, let
us introduce first a multipolar decomposition of the gradient of the
scalar field
\begin{equation}
\Phi^{\ell}_{\alpha}(x^{\mu}) = \nabla^{}_{\alpha}\sum_{m=-\ell}^{\ell}
\Phi^{m}_{\ell}(t,r)\,Y^{m}_{\ell}(\theta,\varphi) \,,\label{l_retarded}
\end{equation}
so that the gradient of the retarded field is given by
\begin{equation}
 \Phi^{}_{\alpha}(x^{\mu})=\sum_{\ell=0}^{\infty}\Phi^{\ell}_{\alpha}(x^{\mu})\,.
\end{equation}
Obviously, $\Phi^{}_{\alpha}(x^{\mu})$ also diverges at the particle worldline
although the $\Phi^{\ell}_{\alpha}$ are finite.  
Following the discussion of the previous section, the gradient of $\Phi$,
and hence the self-force, can be regularized by splitting the full retarded
field into a singular part and a regular part (see, for instance,~\cite{Detweiler:2002mi,Poisson:2004lr}), 
such that they satisfy equations (\ref{eqsingular})-(\ref{regparticlemotion}).  That is, the
regular part of the gradient of the scalar field will be given, on the
particle location, by
\begin{equation}
\Phi^{R}_{\alpha}(\textit{z}^{\mu}(\tau))= \mathop{\lim}
\limits_{x^{\mu}\to \textit{z}^{\mu}(\tau)}\sum_{\ell=0}^{\infty}
\left(\Phi^\ell_{\alpha}(x^{\mu})-{\Phi^{S,\ell}_{\alpha}}(x^{\mu})
\right) \,.
\end{equation}
The singular field $\Phi^{S}$ is known in a neighborhood of the particle 
worldline.  In particular, the multipoles of the singular part of the
gradient of the scalar field at the particle worldline are given 
by~\cite{Barack:1999wf,Barack:2000eh,Barack:2001bw,Barack:2001gx,Barack:2002mha,Barack:2002bt,Mino:2001mq}:
\begin{eqnarray}
\lim_{x^{\mu}\to z^{\mu}(\tau)} \Phi^{S,\ell}_{\alpha} & = & q\left[ 
\left(\ell+\frac{1}{2}\right) A^{}_{\alpha} + B^{}_{\alpha} +
\frac{C^{}_{\alpha}}{\ell +\frac{1}{2}} \right. \nonumber \\
&-& \left. \frac{2\sqrt{2}D^{}_{\alpha}}{(2\ell-1)(2\ell+3)}+... \right]\,, 
\label{l_sing}
\end{eqnarray}
where $A^{}_{\alpha}$, $B^{}_{\alpha}$, $C^{}_{\alpha}$, $D^{}_{\alpha}$, \ldots\, are 
called the regularization 
parameters~\cite{Barack:2001gx,Barack:2002mha,Barack:2002bt,Mino:1997nk}.  They are independent of $\ell$, 
but depend on the particle dynamics.  The singular part corresponds to the
first three terms which lead to quadratic, linear, and logarithmical divergences
when we sum over $\ell$. 
The remaining terms form a convergent series that does not contribute to the
self-force (each of them).   In our approximate calculations, we maintain the 
$D^{}_{\alpha}$ term as it accelerates the convergence of the series as we increase 
the number of multipoles included~\cite{Detweiler:2002gi,Haas:2006ne}.  
For the case of interest of this work, circular geodesics, the non-vanishing 
regularization parameters are $A^{}_{r}$, $B^{}_{r}$, 
and $D^{}_{r}$~\cite{Barack:2002mha,Barack:1999wf,Detweiler:2002gi}, which can
be written as follows:
\begin{eqnarray}
A^{}_{r} = -\frac{\sigma^{}_p}{r^{2}_{p}}\frac{\sqrt{1-3M/r^{}_p}}{1-2M/r^{}_{p}}
\,, \label{Ar} 
\end{eqnarray}
\begin{eqnarray}
B^{}_{r} = -\frac{1}{r^{2}_{p}}\sqrt{\frac{1-3M/r^{}_p}{1-2M/r^{}_{p}}}
\left[F^{}_{1/2}-\frac{1-3M/r^{}_{p}}{2(1-2M/r^{}_{p})}F^{}_{3/2} \right]\,, 
\label{Br}
\end{eqnarray}               
\begin{eqnarray}             
D^{}_{r}&=& \frac{1}{r^{2}_{p}}\sqrt{\frac{2(1-2M/r^{}_{p})}{1-3M/r^{}_p}}\left\{ 
-\frac{M}{2r^{}_{p}}\frac{1-2M/r^{}_{p}}{1-3M/r^{}_{p}}F^{}_{-1/2} \right. 
\nonumber \\ 
&-& \frac{(1-M/r^{}_{p})(1-4M/r^{}_{p})}{8(1-2M/r^{}_{p})} F^{}_{1/2} \nn \\ 
&+& \frac{(1-3M/r^{}_{p})(5-7M/r^{}_{p}-14M^2/r^{2}_{p})}{16(1-2M/r^{}_{p})^2}F^{}_{3/2} 
\nonumber \\
&-& \left. \frac{3(1-3M/r^{}_{p})^2(1+M/r^{}_{p})}{16(1-2M/r^{}_{p})^2}F^{}_{5/2}\right\} 
\,, \label{Dr}
\end{eqnarray}
where $\sigma^{}_p$ is a sign that takes the value $+1$ when the limit in 
equation~(\ref{l_sing}) is performed from the {\em right} ($r\geq r^{}_{p}$)
and $-1$ when the limit is performed from the {\em left} ($r\leq r^{}_{p}$).
The coefficients $F^{}_{Q}$ can be computed in terms of hypergeometric functions
as follows (see~\cite{Abramowitz:1970as}, equation~(15.1.1)):
\begin{equation}
F^{}_Q = {}^{}_2F^{}_1\left(Q,\frac{1}{2};1;\frac{M}{r^{}_{p}f(r^{}_{p})}\right) \,.
\end{equation}
Since the only non-vanishing component of the regularization coefficients is the
radial one, this is the only component of the self-force that is actually 
singular, and hence the only one to be regularized.  In this way the 
regularized self-force is:
\begin{equation}
 F^R_{\alpha}= q\, \Phi^{R}_{\alpha}(z^{\mu}(\tau))\,.
\end{equation}

\section{A New Time-Domain Framework for Simulations of EMRI{s}}\label{timedomainframework}

In this section we introduce all the ingredients of a new 
time-domain method to evolve the equation(s) for a scalar
field generated by a charged particle orbiting a nonrotating
black hole, and also to compute the self-force from the
evolved solution.

\subsection{Mathematical Preliminaries}
\label{mathematicalstuff}
We are going to introduce some mathematical developments needed in order to 
compute, using numerical algorithms based on the PSC method, the self-force on 
a charged particle on circular geodesics.  We discuss first
the particular formulation that we implement numerically.    
Since our numerical framework deals with multiple domains it is
useful to adopt a first-order formulation of the scalar field equation. 
The main reason is that first-order formulations of PDEs can be adapted to
the hyperbolic character of the underlying equation, and hence they are
suitable for a correct communication between domains.

We perform a first-order reduction of (\ref{master}) by introducing the 
following new variables
\begin{eqnarray}
\psi^{m}_{\ell}(t,r) & = & r\,\Phi^{m}_{\ell}(t,r)\,, \label{psi}\\
\phi^{m}_{\ell}(t,r) & = & \partial^{}_t\psi^{m}_{\ell}(t,r)\,, \label{phi}\\
\varphi^{m}_{\ell}(t,r) & = & \partial^{}_{\rsu}\psi^{m}_{\ell}(t,r)\,. 
\label{varphi}
\end{eqnarray}
The evolution equations for $\mb{U} = (\psi^{m}_{\ell},\phi^{m}_{\ell},\varphi^{m}_{\ell})$,
that is for a given harmonic $(\ell,m)$, 
written in matrix form, are: 
\begin{eqnarray}
\partial^{}_t\mb{U}=\mathbb{A}\cdot\partial^{}_{\rsu}\mb{U}+\mathbb{B}\cdot\mb{U}
+\mb{S}\,, \label{evosystem}
\end{eqnarray}
where
\begin{eqnarray}
\mathbb{A} = \begin{pmatrix}  0 & 0 & 0 \\ 
                              0 & 0 & 1 \\
                              0 & 1 & 0 \\
\end{pmatrix}\,,~~
\mathbb{B} = \begin{pmatrix}  0 & 1 & 0 \\
                   -V^{}_{\ell} & 0 & 0 \\
                              0 & 0 & 0 \\
\end{pmatrix}\,,
\end{eqnarray}
and
\begin{eqnarray}
\mb{S} = \left(0,-\frac{S^{m}_{\ell}}{f(r^{}_{p})}\delta(\rsu-\rsu_{p}(t)),0\right)\,.
\end{eqnarray}
It can be seen that~(\ref{evosystem}) is a first-order symmetric hyperbolic
system of PDEs, that is, it has the maximum degree of hyperbolicity, as expected
of a system that comes from the reduction of a wave-type equation.  The characteristic structure is 
described by the matrix $\mathbb{A}$.

Given that discontinuities in the solution are only allowed across characteristics,
and given that due to the singular character of the source some of our variables
will have jumps across the radial particle location, it is important to study
these jumps using the formulation just introduced.  To approach this question
it is convenient to divide the spatial domain (the radial direction as parametrized
by the coordinate $\rsu$) into two disjoint regions, one to the left of the particle 
($\rsu<\rsu_{p}(t)$), and one to the right of the particle 
($\rsu>\rsu_{p}(t)$).   Then, we can write the solution of the equations~(\ref{evosystem})
in the form~(see also~\cite{Sopuerta:2005gz}):
\begin{eqnarray}
\mb{U} = \mb{U}^{}_{-}(t,r)\Theta(\rsu_{p}(t)-\rsu) + \mb{U}^{}_{+}(t,r)
\Theta(\rsu - \rsu_{p}(t))\,, \label{globalsolution}
\end{eqnarray}
where $\Theta$ denotes the Heaviside step function. 
Introducing~(\ref{globalsolution}) into~(\ref{evosystem}) we can derive the jump
across the particle location, defined as
\begin{eqnarray}
\salto{\lambda} = \mathop{\lim }\limits_{\rsu \to \rsu_{p}}\lambda^{}_{+}(t, \rsu)
-\mathop{\lim }\limits_{\rsu \to \rsu_{p}}\lambda^{}_{-}(t, \rsu)\,,
\end{eqnarray}
for the different variables of our problem.  We find the following result:
\begin{eqnarray}
\salto{\psi^{m}_{\ell}} = 0\,,~~ 
\salto{\phi^{m}_{\ell}} = 0\,,~~
\salto{\varphi^{m}_{\ell}} =  \frac{S^{m}_{\ell}}{f(r_p)}\,. \label{jump}
\end{eqnarray}
These are the conditions we have to impose in our numerical evolution 
algorithm.  In practice, these conditions have to be imposed on the
{\em characteristic} fields, the fields associated to the characteristics.
In our problem there are two different types of characteristics:
(i) The $\{t=\mbox{const.}\}$ surfaces, and (ii) the null surfaces
$\{t\pm \rsu = \mbox{const.}\}$.  The associated characteristic
fields are $\psi_{\ell}^{m}$, and $\phi^{m}_{\ell}\mp \varphi^{m}_{\ell}$,
respectively.

Finally, our system of equations has to be complemented with initial 
conditions and boundary conditions.  The initial conditions, since
we are dealing with a first-order system of PDEs, consist in prescribing
$\mb{U}^{}_{o}(\rsu) = \mb{U}(t=t^{}_{o},\rsu)\,$.
We need boundary conditions at the ends of the spatial domain.  The physical
spatial domain is $\rsu\in (-\infty,+\infty)$, with $\rsu\rightarrow
-\infty$ corresponds to the horizon location, while $\rsu\rightarrow
+\infty$ corresponds to spatial infinity.  In our numerical computations we
will take a {\em truncated} domain: $\rsu\in [\rsu_\Hor,\rsu_\Inf]$,
and therefore we need to prescribe outgoing boundary conditions at the ends
of that domain (also called {\em absorbing} boundary conditions).  As an
approximation we take the conditions that are exact for the case in which
there is no potential, that is 
\begin{eqnarray}
&& \phi^{m}_{\ell}(t,\rsu_{\Hor}) - \varphi^{m}_{\ell}(t,\rsu_{\Hor})
= 0\,,~~ \\[2mm]
&& \phi^{m}_{\ell}(t,\rsu_{\Inf}) + \varphi^{m}_{\ell}(t,\rsu_{\Inf})
= 0\,. \label{outgoingbcs}
\end{eqnarray}
This is the leading approximation for the outgoing boundary conditions.
They can be improved by analyzing the solution near $\rsu\rightarrow
\pm\infty$.  However, we can always take values of $(\rsu_{\Hor},
\rsu_{\Inf})$ such that the boundaries are not in causal contact with
the particle location, avoiding contamination of the solution due to
propagation of unphysical modes from the boundaries.

\subsection{Using the PseudoSpectral Collocation Method}
\label{pscalgorithm}

We now describe in some detail the numerical techniques we use to solve
for the full retarded scalar field and to compute from it the regularized
self-force.

\begin{figure}[htp]
\centering
\includegraphics[width=0.5\textwidth]{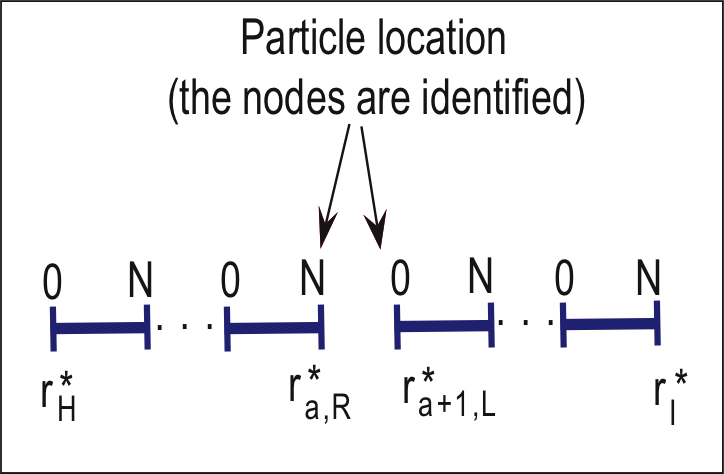}
\caption{The figure shows the structure of the one-dimensional spatial grid,
the division in subdomains and the location of the particle at the interface
between two of them.  The boundaries of the domain have coordinates 
$\rsu_{\Hor}$ (this coordinate will have a finite value and is meant to
approach the BH horizon at $\rsu =-\infty$) and $\rsu_{\Inf}$ (this coordinate
will have too a finite value and is meant to approach spatial infinity at 
$\rsu = +\infty$).  The particle is at the boundaries of two different boundaries
(which are identified), which have coordinates $\rsu_{a,R}$ and $\rsu_{a+1,L}$
that satisfy the relation~(\ref{particleatinterface}).}\label{multidomain}
\end{figure}

In order to solve for the PDEs that describe the scalar field, 
equations~(\ref{evosystem}), we use the PSC to discretize in space.
Once this is done we obtain a system of ordinary differential equations 
(ODEs) that can be solve by
using the method of lines (see, e.g.~\cite{Gustafsson:1995tb}) applying a 
convenient ODE solver.
In general, spectral methods can approximate solutions of PDEs by
expanding the variables in a given basis of functions, and then by using 
an appropriate criterium that forces this expansion to approach the exact
solution as we increase the number of functions included in the expansion.
In the case of the PSC method, the criterium consist in imposing the
solution exactly at a set of {\em collocation} points (see, e.g.~\cite{Boyd}, 
for details on the PSC method).   In our work, we use the Chebyshev 
polynomials, $\{T^{}_{n}(X)\}$ ($X \in [-1,1]$), as the basis functions 
(see Appendix~\ref{chebyshev} for definitions), and for the collocation points 
we take a {\em Lobatto-Chebyshev} grid.   If we want to use a grid of $N$
collocation points, the Lobatto-Chebyshev grid is made out of the zeros of
the following polynomial~(see Appendix~\ref{chebyshev}):
\begin{eqnarray}
 (1-X^{2})T'_{N}(X) = 0\,,
\end{eqnarray}
where the prime indicates differentiation with respect to $X$.
The zeros can be written as follows:
\begin{eqnarray}
 X^{}_i = -\cos\left(\frac{\pi\,i}{N}\right)~~(i=0,1,\ldots,N)\,.
 \label{chebyshevlobattogrid}
\end{eqnarray}
Now, we need to map the domain in which the Chebyshev polynomials are defined
(we will call it the {\em spectral domain}) to the physical domain.  Before
doing that, we have to mention that for computational reasons that we
discuss later, we split the computational domain, $\Omega=
\left[\rsu_H,\rsu_I\right]$, into a number of subdomains ($D$): 
\begin{equation}
\Omega = \bigcup^{D}_{a=1} \Omega^{}_a\,, ~~
\Omega^{}_a = \left[ \rsu_{a,L}, \rsu_{a,R}\right]\,,
\end{equation}
where $\rsu_{a,L}$ and $\rsu_{a,R}$ are the left and right boundaries of the
subdomain $\Omega^{}_{a}$.  They are disjoint subdomains, that is 
$\rsu_{a-1,R}=\rsu_{a,L}$ (see figure~\ref{multidomain} and the caption there).  

We apply the PSC method to each subdomain,
in the sense that our variables have different expansions in Chebyshev polynomials
in each subdomain.  These different expansions are related by using appropriate
boundary conditions that we discuss in section~\ref{evolutionalgorithm}.  
To apply the PSC method to each subdomain, we map the physical subdomain $\Omega_a$ 
($a=1,\ldots,D$) to the spectral domain, $[-1,1]$, using a linear mapping 
$X^{}_{a}\,:\, \left[ \rsu_{a,L}, \rsu_{a,R}\right]
\longrightarrow \left[-1,1\right]\,,$ with
\begin{eqnarray}
\rsu \longrightarrow X^{}_{a}(\rsu)= \frac{2\rsu- \rsu_{a,L}- \rsu_{a,R}}{ \rsu_{a,R}-
\rsu_{a,L} } \,.    \label{map1}
\end{eqnarray} 
The inverse correspondence is another linear mapping, $\left.\rsu\right|^{}_{\Omega^{}_{a}}\,:\, 
\left[-1,1\right] \longrightarrow  \left[ \rsu_{a,L}, \rsu_{a,R}\right]\,,$ with
\begin{eqnarray}
X  \longrightarrow  \left.\rsu(X)\right|^{}_{\Omega^{}_{a}} = 
\frac{\rsu_{a,R}-\rsu_{a,L}}{2}X+\frac{\rsu_{a,L}+ \rsu_{a,R}}{2}\,.
\end{eqnarray}

The variables of our problem are arranged in a vector, $\mb{U}$.  Then,
at a given domain $\Omega_a$ ($a=1,\ldots,D$), we have the following spectral expansion 
of our variables:
\begin{equation}
\mb{U}^{}_{N}(t,\rsu) = \sum_{n=0}^N \mb{a}^{}_n(t)\, T^{}_n(X^{}_a(\rsu)) \,,
\label{spectralrepresentation}
\end{equation}
where the $\mb{a}^{}_n$ are (time-dependent) vectors that contain the spectral coefficients 
of the expansion of our variables.  In the PSC method, we have also a {\em physical}
expansion, which looks as follows:
\begin{equation}
\mb{U}^{}_{N}(t,\rsu) = \sum_{i=0}^N \mb{U}^{}_i(t)\, {\cal C}^{}_i(X^{}_a(\rsu))\,,
\label{physicalrepresentation}
\end{equation}
where ${\cal C}^{}_i(X)$ are the {\em cardinal} functions~\cite{Boyd} associated with
our choice of basis functions (Chebyshev polynomials) and set of collocation points
(Lobatto-Chebyshev grid) [see Appendix~\ref{chebyshev} for details].  The cardinal 
functions have the following remarkable property:
\begin{equation}
{\cal C}^{}_i(X^{}_{j}) = \delta^{}_{ij} \,.
\end{equation}
In this way, the time-dependent (vector) coefficients, $\{\mb{U}^{}_{i}\}$, of the 
expansion~(\ref{physicalrepresentation}) are the values of our variables at the 
collocation points
\begin{equation}
\mb{U}(t,\rsu(X^{}_{i})) = \mb{U}^{}_{i}(t)\,.
\end{equation}
These are the variables that one looks for in the PSC method.  
The spectral~(\ref{spectralrepresentation}) and physical~(\ref{physicalrepresentation})
representations are related via a matrix transformation (see Appendix~\ref{chebyshev}).
The computations (float-point operations) required to change representation scale, with 
the number of collocation points, as $N^{2}$, as it can be deduced from the fact that 
it is a matrix transformation.  However, using the change of spectral coordinate given in 
equation~(\ref{changespectralcoord}), we can perform the change of representation by means 
of a discrete Fourier transform using a Fast-Fourier Transform (FFT) algorithm.  
In our numerical codes, we use the routines of the FFTW library~\cite{fftw:2005}.  
Then, the number of computations required for a change of representations scales as 
$\sim N\ln N$ with the number of collocation points.

Changing between representations is useful in order to perform some operations.  
For instance, differentiation is easier in the spectral representation, so we can
transform from the physical to the spectral representation, compute derivatives 
there, and finally transform back to the physical representation.   In the
case of a Chebyshev PSC method, the differentiation process can be described by
the following scheme
\begin{eqnarray}
\partial^{}_{\rsu} :\,\{\mb{U}^{}_i\} \;\stackrel{FFT}{\longrightarrow}\; 
\{\mb{a}^{}_n\} \;\stackrel{\partial^{}_{\rsu}}{\longrightarrow}\;
\{\mb{b}^{}_n\} \;\stackrel{FFT}{\longrightarrow}\;
\{(\partial^{}_{\rsu}{\mb{U}})^{}_i\}\,, \label{pscdifferentiation}
\end{eqnarray}
where $\{\mb{b}^{}_{n}\}$ are the spectral coefficients associated with the
spatial derivative, and their relation to the coefficients of the variables, 
$\{\mb{a}^{}_{n}\}$, is given by (see, e.g.~\cite{Boyd})
\begin{eqnarray}
&& \mb{b}^{}_N = \mb{b}^{}_{N-1}=0\,, \\
&& \mb{b}^{}_{n-1} = \frac{1}{c^{}_n}\left\lbrace 2n\mb{a}^{}_n+\mb{b}^{}_{n+1}
\right\rbrace~(n=N-1,\ldots,1)\,,
\end{eqnarray}
and
\begin{eqnarray}
{c}^{}_n=\left\{\begin{array}{ll} 2  & \text{for}~n=0\,, \\[2mm]
                                  1  & \text{otherwise\,.}
\end{array}\right. \label{cns}
\end{eqnarray}

In the PSC method, we find a discretization of our system of equations~(\ref{evosystem})
by imposing them at every collocation point.  In practice, this is done by introducing
the expansion~(\ref{physicalrepresentation}) into the equations~(\ref{evosystem}), and then 
we evaluate the result at every collocation point of our Chebyshev-Lobatto 
grid~(\ref{chebyshevlobattogrid}).  We obtain a system of ODEs for the variables
$\{\mb{U}^{}_{i}(t)\}$
\begin{eqnarray}
\dot{\mb{U}}^{}_i = \mathbb{A}\cdot (\partial^{}_{\rsu}\mb{U})^{}_{i} 
+\mathbb{B}\cdot\mb{U}^{}_i+\mb{S}^{}_i \,, \label{odes}
\end{eqnarray}
where the dot denotes differentiation with respect to the time coordinate $t$,
and $(\partial^{}_{\rsu}\mb{U})^{}_{i}$ has to be interpreted according to the 
scheme in equation~(\ref{pscdifferentiation}).

\subsection{Evolution Algorithm}
\label{evolutionalgorithm}
In this section we discuss the details of the time evolution of the discrete equations
we derived in the previous section.  In particular, we describe how to introduce the
particle in the multi-domain PSC method that we propose.  The main argument is the
following: If we put the particle inside one of the subdomains $\Omega^{}_{a}$ we
are introducing in the equations~(\ref{evosystem}) a singular term that, in the case
of a scalar charged particle, will produce solutions that are not differentiable 
(in the sense that the derivative with respect to $\rsu$ is not single-valued at the 
particle location).  
That is, our solution will not be smooth and hence we cannot expect the PSC method
to converge exponentially to the true solution.  That would spoil one of the main
motivations for using this numerical technique, that is, the accuracy that the PSC
method provides.  To avoid this (and this is part of the reason for using a multi-domain
scheme) we put the particle at the interface between two subdomains.  If the subdomains
are $\Omega^{}_{a}$ and $\Omega^{}_{a+1}$, we have:
\begin{equation}
\rsu_{p} = \rsu_{a,R} = \rsu_{a+1,L}\,. \label{particleatinterface}
\end{equation}
Since we have restricted ourselves to the case of circular orbits, once the grid
has been set we do not need to change it during the evolution.  In the case of 
generic orbits, in order to maintain the particle at the interface between two
subdomains we have either to make a coordinate change or implement a moving grid
scheme (a similar situation was confronted in~\cite{Sopuerta:2005gz}).  
We discuss this further in section~\ref{discussion}. 

For each subdomain $\Omega^{}_{a}$ ($a=1,\ldots,D$), we evolve the set of 
ODEs~(\ref{odes}) independently.  Since we are locating the particle at the
interface between two subdomains, the last term in~(\ref{odes}), $\mb{S}$, 
does not appear. This term is the source term that accounts for the particle's
energy density [see equation~(\ref{source})].  The main implication of this 
setup is that we have to solve the homogeneous field equations, which are  
equations with smooth solutions  and hence, the advantages of the PSC method are
preserved.  Then, in our framework, the contributions
of the particle to the solution appear as boundary conditions between the 
subdomains.  In summary, we evolve the (homogeneous) field equations for each subdomain 
independently and connect their solutions by boundary conditions at the interfaces. 
The equations are evolved using a Runge-Kutta (RK) solver 
(see~\cite{Butcher:2008jb,Press:1992nr}), typically a RK4 algorithm.

The key point in the evolution is the imposition of boundary conditions at the
interfaces between subdomains.  There are two possible situations for a given
interface: (i) The particle is not there.  In this case we only have to impose
the continuity of the solution [that would correspond to imposing the junction
conditions given in equation~(\ref{jump}) with zero right-hand sides].  
(ii) The particle is there.  In this case we have to impose the junction conditions
in equation~(\ref{jump}) where $S^{m}_{\ell}$ is given in equation~(\ref{source}).
The question is then how to impose these boundary conditions numerically within
our framework.  There are different ways of doing this, and in this work we
have chosen to impose this boundary conditions at the interfaces between subdomains
in a dynamical way, via {\em penalty} terms.

The penalty method is a well-known technique and has been applied to several numerical
schemes to solve PDEs (for the PSC method, see~\cite{Hesthaven:2000jh} and references
therein).  It is relatively simple to implement it for elliptic-type problems, that is,
for problems that do not require evolution {\em in time}.  For time-dependent problems,
it can also be implemented but the result is not always numerically stable
In our framework, the implementation of the penalty method goes as follows: First, 
let us consider the two domains around the particle's location, say 
$\Omega^{}_{a}$ and $\Omega^{}_{a+1}$, so that equation~(\ref{particleatinterface}) holds.
Then, let us consider the solutions at these two subdomains, $\mb{U}^{}_{a}$ and
$\mb{U}^{}_{a+1}$.  The equations for the inner points are just equations~(\ref{odes})
(with $\mb{S}^{}_i = 0$),
and the equations for the nodes at the interface between these subdomains, $\rsu_{a,R}$
and $\rsu_{a+1,L}$ (which are identified), are modified 
in the following way:
For the subdomain
$\Omega^{}_{a}$ (we have simplified the notation by dropping the harmonic indices $\ell$
and $m$)
\begin{eqnarray}
\partial^{}_t\psi^{}_{a,R} & = & \phi^{}_{a,R} - \tau^{a,R}_{\psi}\left[\psi^{}_{a,R} 
-\psi^{}_{a+1,L}\right]\,, \label{penaltyR1}\\
\partial^{}_t \phi^{}_{a,R} & = & \partial^{}_{\rsu} \varphi^{}_{a,R} - V^{}_{p}\,\psi^{}_{a,R} 
-\frac{\tau^{a,R}_{\phi}}{2}\left\{ \phi^{}_{a,R}+\varphi^{}_{a,R}\right.  \nonumber \\ 
& - & \left. (\phi^{}_{a+1,L} + \varphi^{}_{a+1,L}) - \salto{\varphi} \right\}\,, \label{penaltyR2} \\
\partial^{}_t\varphi^{}_{a,R} & = & \partial^{}_{\rsu}\phi^{}_{a,R} 
-\frac{\tau^{a,R}_{\varphi}}{2}\left\{\phi^{}_{a,R} + \varphi^{}_{a,R} \right.\nonumber \\
& - & \left. (\phi^{}_{a+1,L} + \varphi^{}_{a+1,L}) - \salto{\varphi} \right\}\,, \label{penaltyR3}
\end{eqnarray}
and for the subdomain $\Omega^{}_{a+1}$
\begin{eqnarray}
\partial^{}_t\psi^{}_{a+1,L} & = & \phi^{}_{a+1,L} - \tau^{a+1,L}_{\psi}\left[\psi^{}_{a+1,L}
-\psi^{}_{a,R}\right]\,, \label{penaltyL1} \\
\partial^{}_t\phi^{}_{a+1,L} & = & \partial^{}_{\rsu}\varphi^{}_{a+1,L} - V^{}_{p}\,
\psi^{}_{a+1,L}-\frac{\tau^{a+1,L}_{\phi}}{2}\left\{ \phi^{}_{a+1,L}\right. \nonumber \\
& - &\left. \varphi^{}_{a+1,L} - (\phi^{}_{a,R} - \varphi^{}_{a,R}) - \salto{\varphi} \right\}\,, \label{penaltyL2} \\
\partial^{}_t\varphi^{}_{a+1,L} & = & \partial^{}_{\rsu}\phi^{}_{a+1,L} 
-\frac{\tau^{a+1,L}_{\varphi}}{2}\left\{\phi^{}_{a+1,L} -\varphi^{}_{a+1,L}\right. \nonumber \\
& - &\left. (\phi^{}_{a,R} - \varphi^{}_{a,R})  - \salto{\varphi}  \right\}\,, \label{penaltyL3}
\end{eqnarray}
where $V^{}_p = V(r^{}_p)$, the quantity $\salto{\varphi}$ is the analytic jump imposed by the
dynamics and given by equation~(\ref{jump}),  and
\begin{eqnarray}
\psi^{}_{a,R}(t) = \psi(t,\rsu_{a,R})\,,& &\psi^{}_{a+1,L}(t) = \psi(t,\rsu_{a+1,L})\,, \\
\phi^{}_{a,R}(t) = \phi(t,\rsu_{a,R})\,,& &\phi^{}_{a+1,L}(t) = \phi(t,\rsu_{a+1,L})\,, \\
\varphi^{}_{a,R}(t) = \varphi(t,\rsu_{a,R})\,,& &\varphi^{}_{a+1,L}(t) = 
\varphi(t,\rsu_{a+1,L})\,,
\end{eqnarray}
and where $\tau^{a,R}_{\psi,\phi,\varphi}$ and $\tau^{a+1,L}_{\psi,\phi,\varphi}$ are 
(constant) penalty parameters.   The structure of the penalty terms in 
equations~(\ref{penaltyR1})-(\ref{penaltyL3}) obeys the following rationale:
The main idea behind of the penalty terms is to drive the dynamical system 
to satisfy a set of conditions that are not part of the original evolution
equations (like constraints on the variables that have to be satisfied for all
times or boundary conditions), and the strength of the driving (penalty) terms
is controlled by the penalty parameters $\tau^{a,R}_{\psi,\phi,\varphi}$ and 
$\tau^{a+1,L}_{\psi,\phi,\varphi}$.  In our case, in section~\ref{mathematicalstuff}, 
we mentioned the fact that the junction conditions~(\ref{jump}) have to be imposed 
on the characteristic field of our system of PDEs, and these characteristic fields and 
their associated characteristic surfaces where given there.  Therefore, we have
constructed the penalty terms so that the junction conditions that are satisfied 
are those corresponding to the characteristic fields, that is [restoring the $(\ell,m)$ indices]:
\begin{eqnarray}
\salto{\psi^{m}_{\ell}} = 0\,,~~ 
\salto{\phi^{m}_{\ell}\pm \varphi^{m}_{\ell}} = \pm\, \frac{S^{m}_{\ell}}{f(r_p)}\,.
\label{characteristicjump}
\end{eqnarray}
On the other hand, $\psi$ can be seen as a {\em subsidiary} variable, in the 
sense that we can first evolve the equations for $\phi$ and $\varphi$ and then, 
use the result to evolve $\psi$.  Then, we can use a different way of imposing 
the continuity of $\psi$, instead of the penalty method we can just replace the
right-hand sides of equations (\ref{penaltyR1}) and (\ref{penaltyL1}) by 
\begin{eqnarray}
\partial^{}_t\psi^{}_{a,R} & = & (\phi^{}_{a,R}+\phi^{}_{a+1,L})/2\,, 
\label{penaltyR1bis}\\
\partial^{}_t\psi^{}_{a+1,L} & = & (\phi^{}_{a,R}+\phi^{}_{a+1,L})/2\,, 
\label{penaltyL1bis} 
\end{eqnarray}
which ensures the continuity of $\psi$ by construction and we have seen
in our numerical experiments that it is numerically stable (see Sec.~\ref{results}).

Regarding the global boundary conditions ({\em near} the horizon, $\rsu
=\rsu_{\Hor}$, and {\em near} spatial infinity, $\rsu=\rsu_{\Inf}$),
we impose them directly at the corresponding nodes, without using the
penalty method (which is another option).   Since we have made a first-order
reduction of our original PDE~(\ref{master}), we have to adapt the
outgoing/absorbing boundary conditions to the set of variables $\mb{U} =
(\psi,\phi,\varphi)$.  Then, the boundary conditions at $\rsu=\rsu_{\Hor}$
are
\begin{eqnarray}
\partial^{}_t\psi^{}_{1,L} & = & \phi^{}_{1,L}\,, \\
\partial^{}_t\phi^{}_{1,L} & = & -\partial^{}_t\varphi^{}_{1,L}\,, \\
\partial^{}_t\varphi^{}_{1,L} & = & \partial^{}_{\rsu} \phi^{}_{1,L}\,,
\end{eqnarray}
and the boundary conditions at $\rsu=\rsu_{\Inf}$ are
\begin{eqnarray}
\partial^{}_t\psi^{}_{D,R} & = & \phi^{}_{D,R}\,, \\ 
\partial^{}_t\phi^{}_{D,R} & = & \partial^{}_t\varphi^{}_{D,R}\,, \\
\partial^{}_t\varphi^{}_{D,R} & = & \partial^{}_{\rsu} \phi^{}_{D,R}\,,
\end{eqnarray}
and we remark that the first subdomain is number $1$ and the last one is number
$D$, the total number of subdomains.

\section{Results from the Simulations}\label{results}

In this section we describe the results from a series of numerical experiments 
that show the potential of the methods just proposed.  To that end, we have
developed a numerical code that implements the techniques previously described.
This code is based on the C language and the use of the
GNU Scientific Library~\cite{Galasi:2006mg}, mainly for calculations with special functions, 
and the FFTW library~\cite{fftw:2005} for performing FFTs.  
In the implementation of the physical domain in the numerical code we have
adopted a comoving tortoise coordinate: $\rsu_c = \rsu-\rsu_p$.  In this
way, the particle location (and we emphasize again that we only deal with
circular orbits) is always $\rsu_c = 0$.

\begin{figure}[htp!]
\centering  
\includegraphics[width=0.5\textwidth]{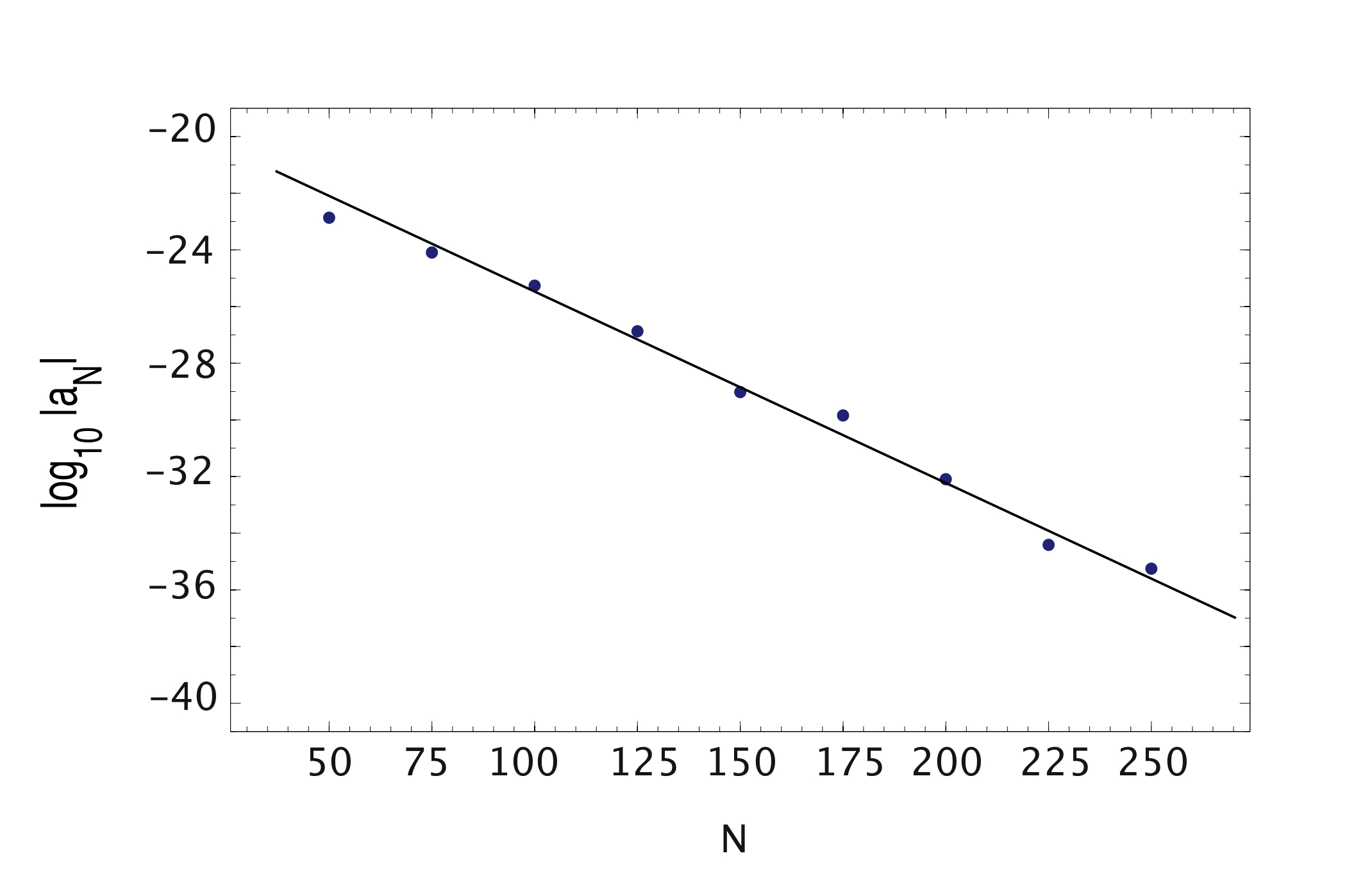}
\caption{This figure shows the dependence of the truncation error (estimated by the logarithm of 
the absolute value of the last spectral coefficient, $\log_{10} |a^{}_N|$, associated with
the field $\psi$)
with respect to the number of collocation points.  This error corresponds to an snapshot of
the evolution of the classical wave equation, for an initial condition given by a moving
Gaussian packet.  The data indicates the exponential convergence of the numerical method.} 
\label{vacuumconvergence}
\end{figure}

The first test we have performed is to study the evolution of a simple 
wave equation using the formulation and methods described in the previous 
sections.  This case corresponds to: $\ell=m=0$, $M=0$ (which implies 
$V^{}_{\ell} = 0$), and $q=0$ (that is, no source, ${\cal S}_\ell^m = 0$).
The test consists in following the propagation of an initial Gaussian packet in a
multidomain grid and to study the convergence of the numerical scheme as the
number of collocation points per subdomain, $N$, increases.  In Figure~\ref{vacuumconvergence},
we show a convergence plot for a simulation that uses two subdomains, 
$\rsu-\rsu_p\in [-550\,M,0]\cup [0,550\,M]$, connected by the penalty method as 
described in subsection~\ref{evolutionalgorithm}.  
The truncation error has been computed in the subdomain where the Gaussian packet is present 
at a chosen time.  As one can see, the truncation error, estimated as the absolute value of 
the last spectral coefficient, $|a^{}_N|$, decreases exponentially with the number of 
collocation points, as expected in the PSC method for smooth solutions.   

The same convergence follows for an initial Gaussian wave packet propagating on a Schwarzschild background,
that is, for the case in which only ${\cal S}_\ell^m = 0$.  This is expected as the mathematical
structure of the equation is essentially the same.  The additional ingredient is the 
excitation of quasinormal modes of the black hole by the initial wave packet.

\begin{figure*}
\centering 
\includegraphics[width=0.33\textwidth]{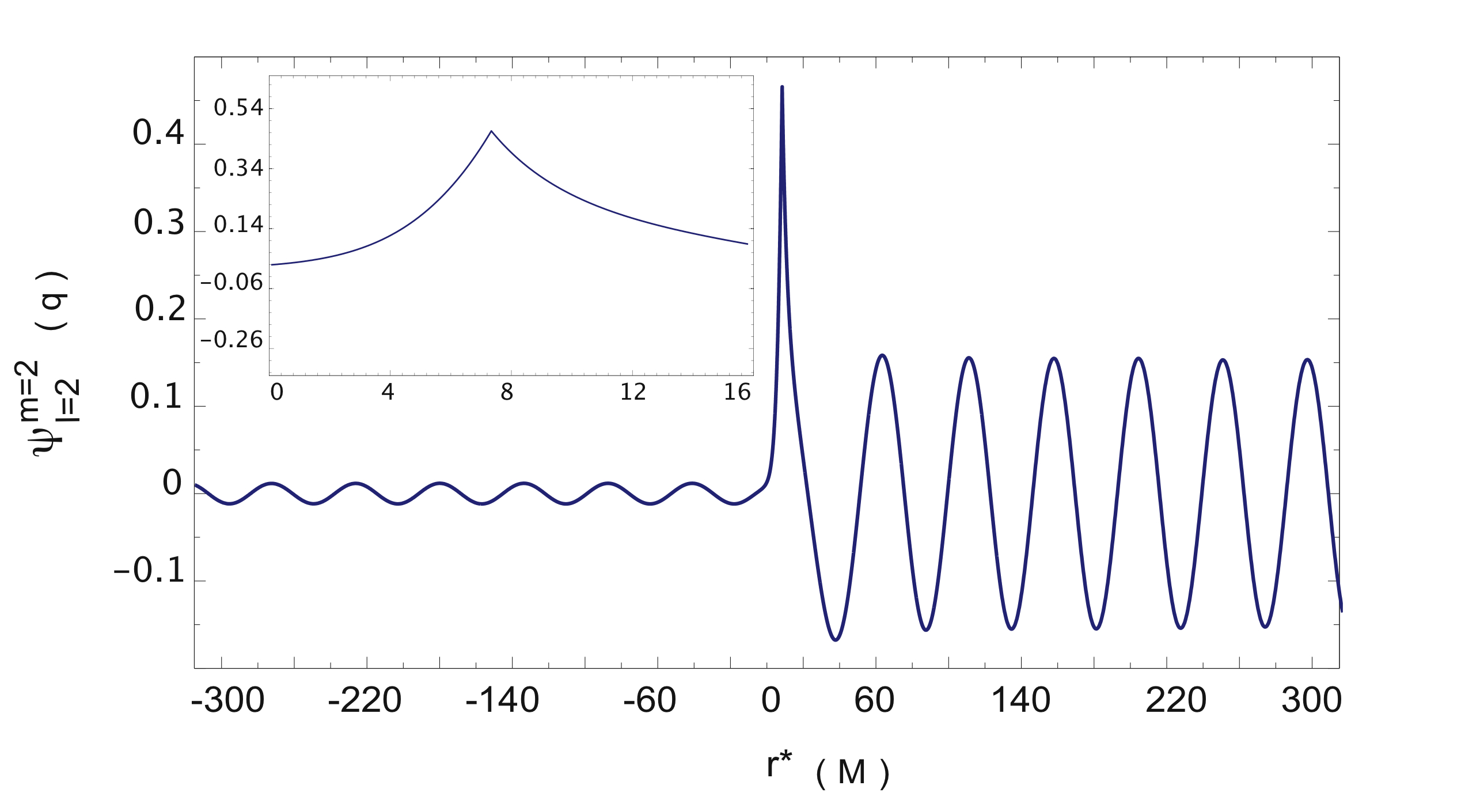}
\includegraphics[width=0.33\textwidth]{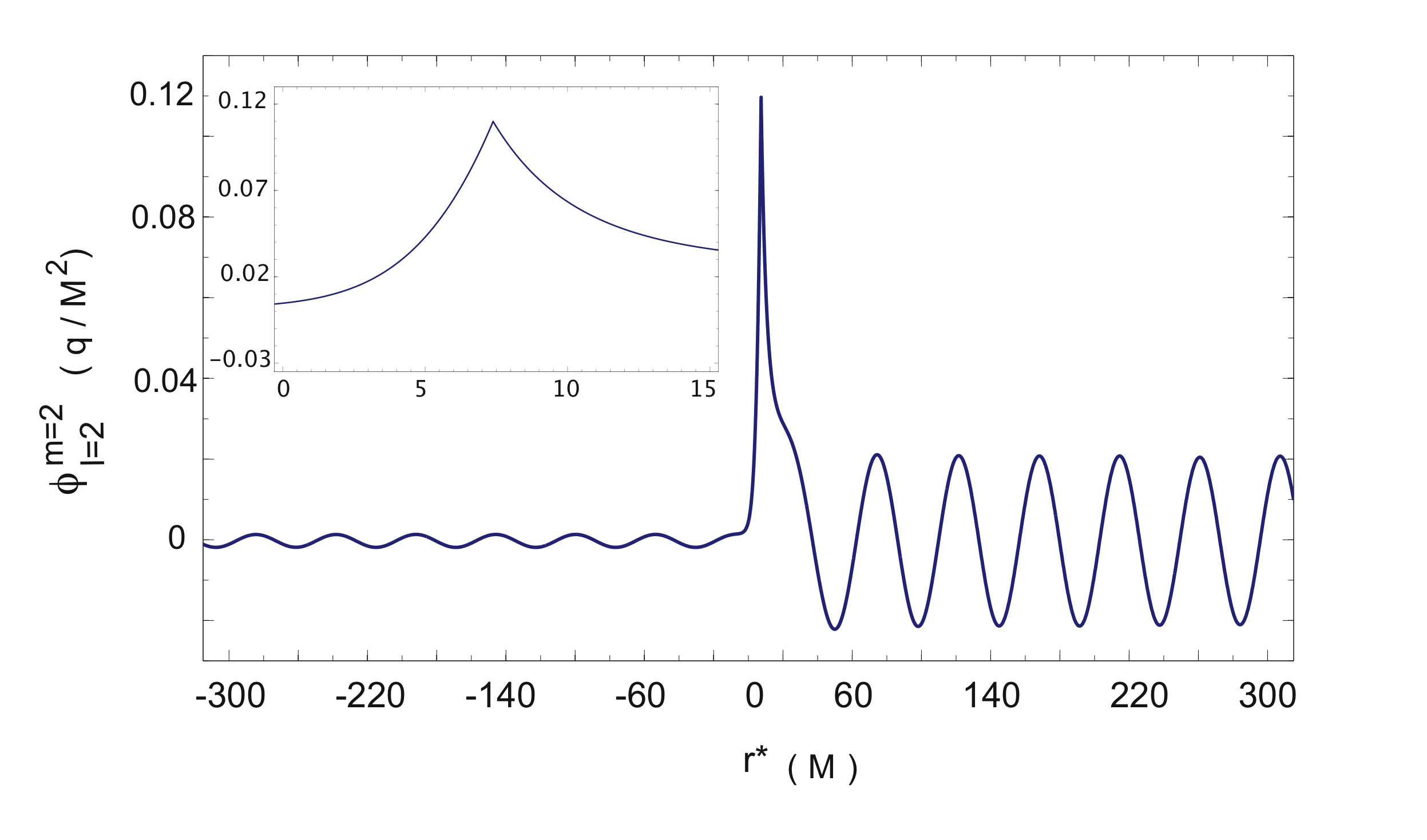}
\includegraphics[width=0.32\textwidth]{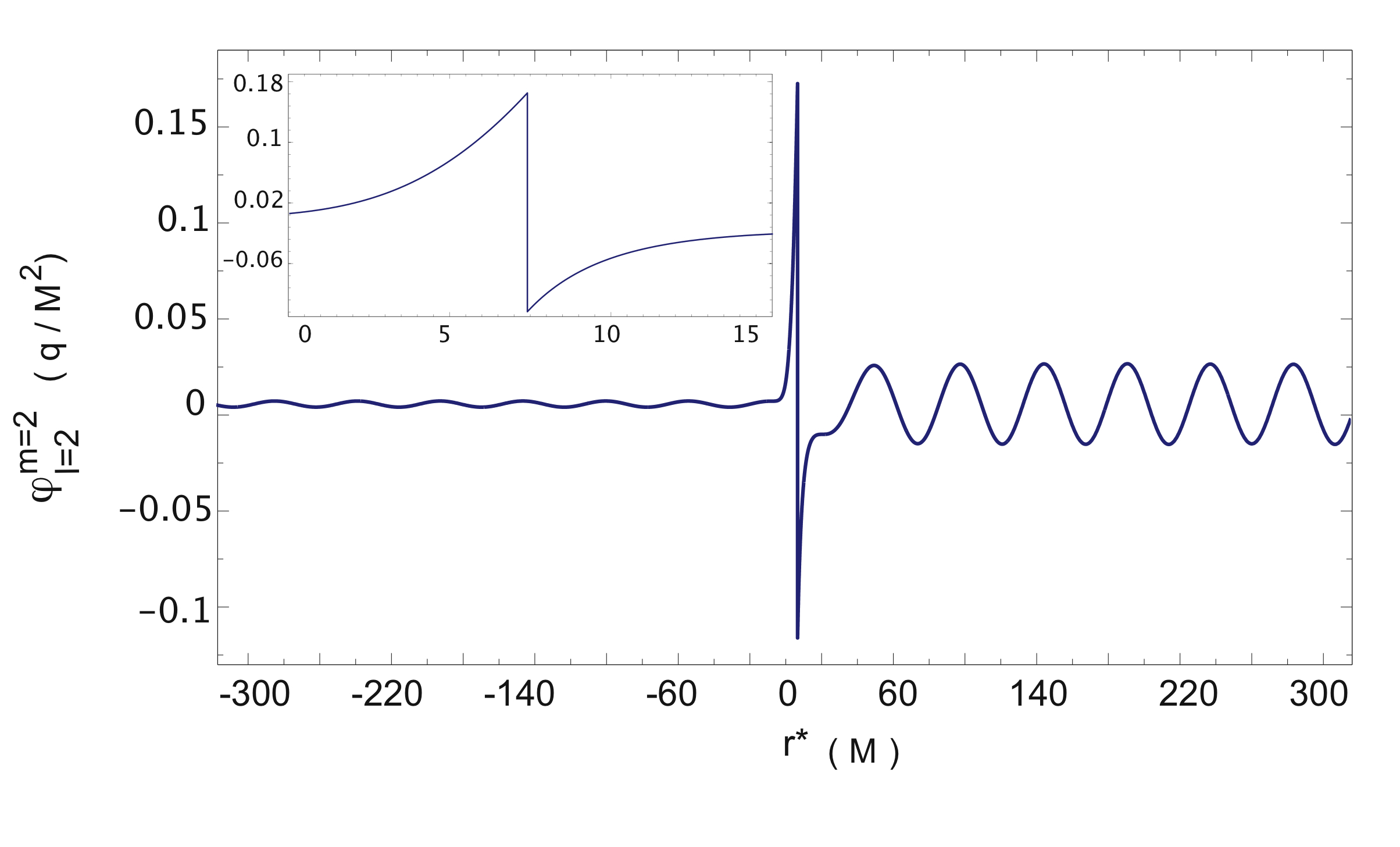}
\caption{We show snapshots of the evolution of the scalar charged particle in circular motion
at the LSO for the mode $\ell=m=2$.   These simulations used $12$ subdomains and $50$ collocation
points per subdomain.  They show the evolution of the variables $\psi_\ell^m$ (left),
$\phi_\ell^m$ (center), and $\varphi_\ell^m$ (right), after a substantial time has passed and a number of
wave cycles have been generated.  In particular they show how the jump in the radial derivative
of the field $\Phi_\ell^m$ is resolved (this information is encoded in the variable $\varphi_\ell^m$)
in this multidomain computational framework.
\label{detailsevolution}}
\end{figure*}

When we introduce the SCO, i.e. the particle, the situation is conceptually different.
If we think in global terms, the presence of the particle implies that the global solution 
(the solution in the whole computational domain) will not be smooth.  Hence, we cannot
expect exponential convergence for the solution of our problem.  However, as we have argued above, our multidomain 
framework avoids the presence of Dirac delta distributions in the equations by locating the particle in the              
interface between subdomains.  Then, the presence of the particle enters through boundary conditions, 
actually matching conditions between subdomains.  Therefore, at each subdomain we will have a 
smooth solution, and hence we expect our numerical solution to converge exponentially
towards it.   In our numerical experiments with a particle (remember we restrict ourselves
to the case of circular orbits), we take {\em zero initial data}, that is
\begin{eqnarray}
 \psi^{m}_\ell(t_o,\rsu) = \phi^{m}_\ell(t_o,\rsu) = \varphi^{m}_\ell(t_o,\rsu)=0\,.
\end{eqnarray}
This initial data is obviously not consistent with Einstein's equations, and as a consequence
the evolution produces an initial unphysical burst.  We have to wait until this unphysical
feature propagates away in order to analyze the solution and obtain physically relevant
results.  In Figure~\ref{detailsevolution} we show snapshots of the evolution of a scalar
charged particle in circular motion at the Last Stable Orbit (LSO), in principle the most demanding 
case (in Schwarzschild the LSO is located at $r^{}_p=6\,M$).  The figure includes details of the 
different variables, $(\psi_\ell^m,\phi_\ell^m,\varphi_\ell^m)$ for $\ell=m=2$, near the particle 
location.  These snapshots illustrate the ability of our method to capture the structure
of the solution near the particle, in particular the ability of resolving the jump in the
radial derivative of the field (snapshot on the right in Figure~\ref{detailsevolution}).

\begin{figure}[htp!]
\centering  
\includegraphics[width=0.5\textwidth]{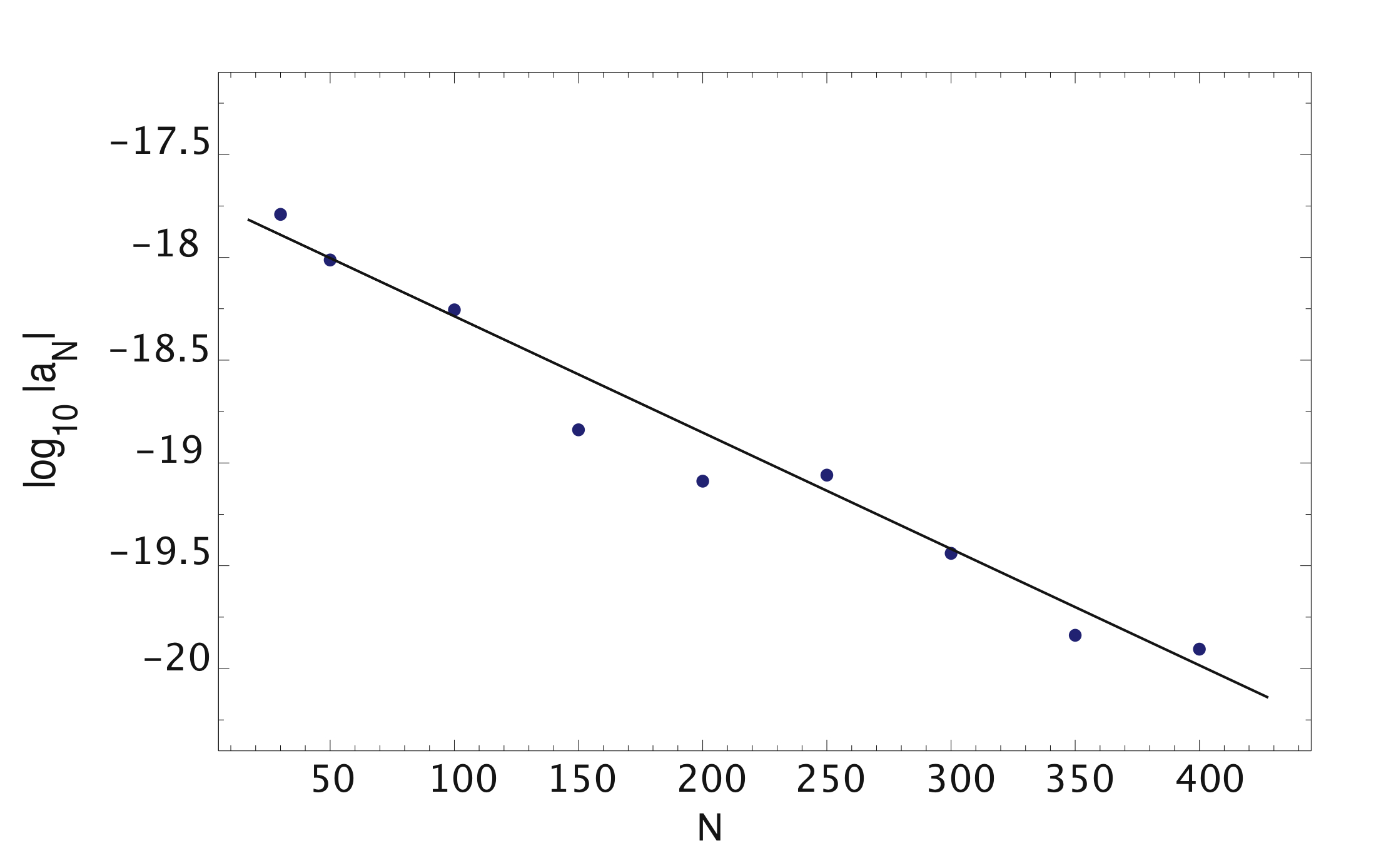}
\caption{This figure shows the dependence of the truncation error ($\log_{10} |a^{}_N|$) with 
respect to the number of collocation points for simulations
of circular motion at the LSO ($r^{}_p=6\,M$) corresponding to the $\ell= m = 2$ mode, in
particular for the field $\psi_2^2$.  The data, taken from the subdomain $\rsu-\rsu_p \in [0,20\,M]$ 
indicates the exponential convergence of the numerical method.
 \label{convergencewithparticle}}
\end{figure}

In order to further validate our numerical code, we have performed simulations with the
particle at the LSO changing the number of collocation points, while leaving the number
of subdomains fixed.  In this way we have checked that our method can achieve
the exponential convergence in each individual subdomain.  In Figure~\ref{convergencewithparticle}
we show a convergence plot obtained from simulations that use four subdomains: 
$\rsu-\rsu_p \in [-550\,M, -20\,M]\cup[-20\,M,0]\cup [0,20\,M]\cup [20\,M,550\,M]$.  The number
of collocation points, $N$, is the same at each subdomain, and in this way we have more resolution
near the particle, where it is most needed.  The figure shows that indeed our numerical scheme
has exponential convergence.

The multidomain feature of our method is useful for another important reason, namely computational
cost.  It is not the same having $N$ points in $D$ domains that having $N\cdot D$ points in one single
domain. This is an important fact to be considered in a situation where the need for resolution comes 
only from some isolated regions of the computational domain.  This is what happens in our problem
and the type of multidomain structure can have dramatic consequences in the calculations.
First of all, for computations in a given time step, the first option involves less calculations.
Second, from the evolution point of view, in the PSC method, the Courant-Friedrichs-Lax (CFL) condition
on the allowed size of the time step, $\Delta t$, is more stringent than in Finite Differences (FD) schemes.
For the PSC method the maximum allowed time step is $\Delta t_{CFL} \sim \pi^2|r^{*}_{R}-r^{*}_{L}|/(4 N^2)$
(this is set by the minimum distance between two collocation points, which occurs at the boundaries of the
subdomains~\cite{Boyd}),
that is, it goes like $1/N^2$ with respect to the number of collocation points, whereas in FD schemes
it goes like $1/N$.  Then, dividing the domain in subdomains can help in having a bigger  $\Delta t_{CFL}$.
In addition, the multidomain scheme can be seen as a way of adaptivity, in the sense that we can construct
small subdomains for the regions that need to be well resolved, essentially near the particle, and large
subdomains for the regions that do not need to have high resolution, essentially far away from the 
particle.

\begin{figure*}
\centering
\includegraphics[width=0.33\textwidth]{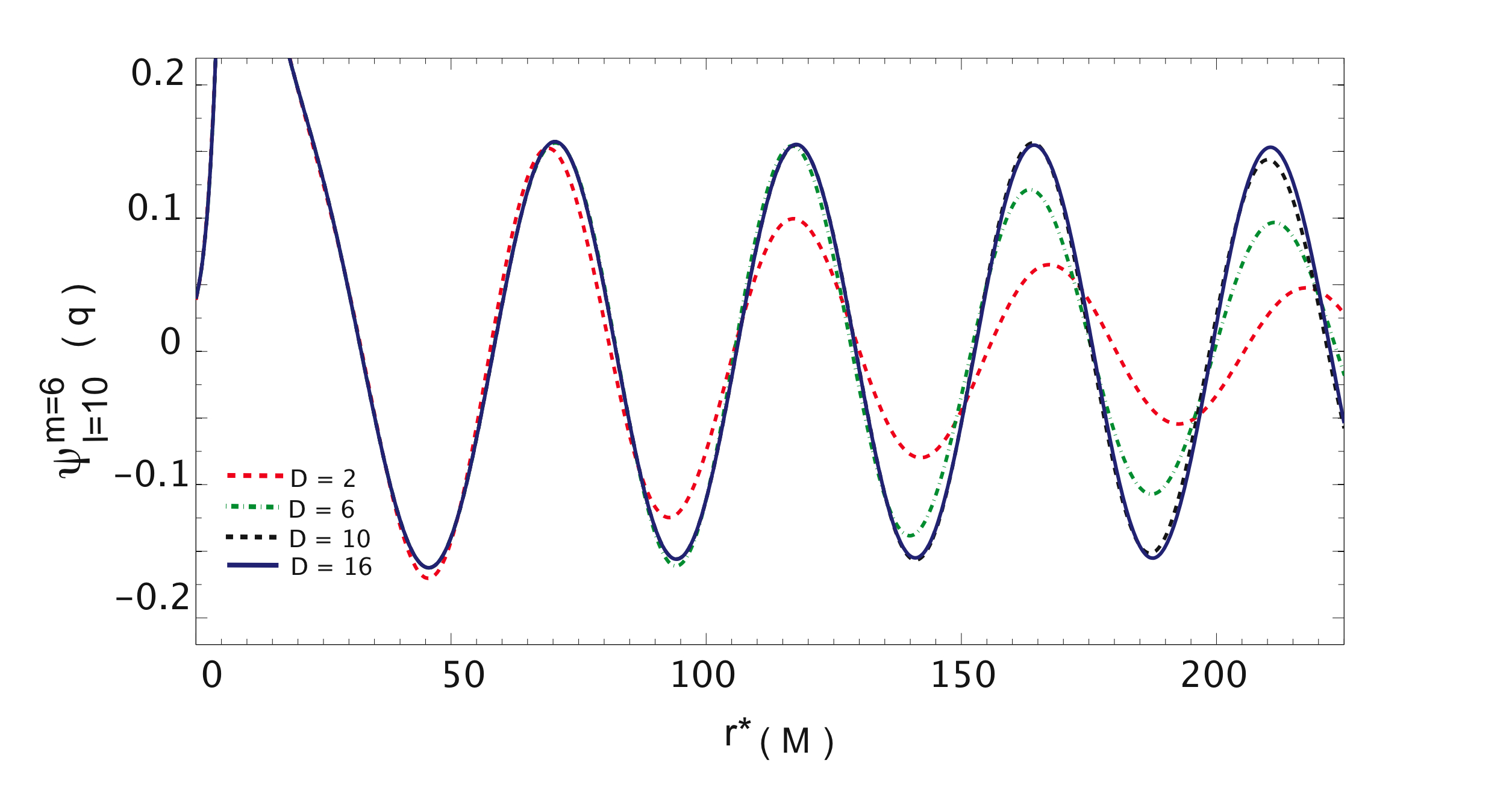}
\includegraphics[width=0.325\textwidth]{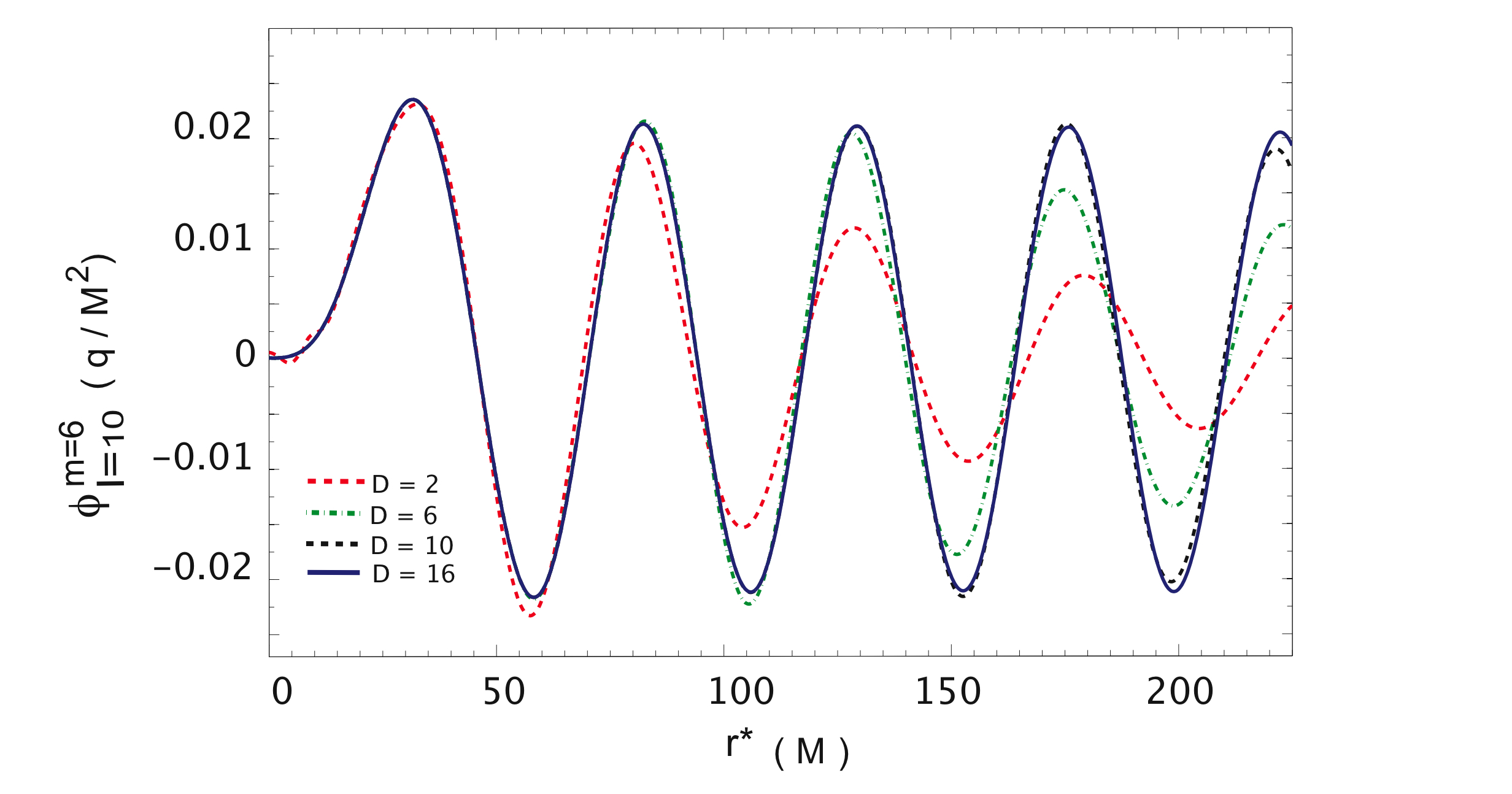}
\includegraphics[width=0.33\textwidth]{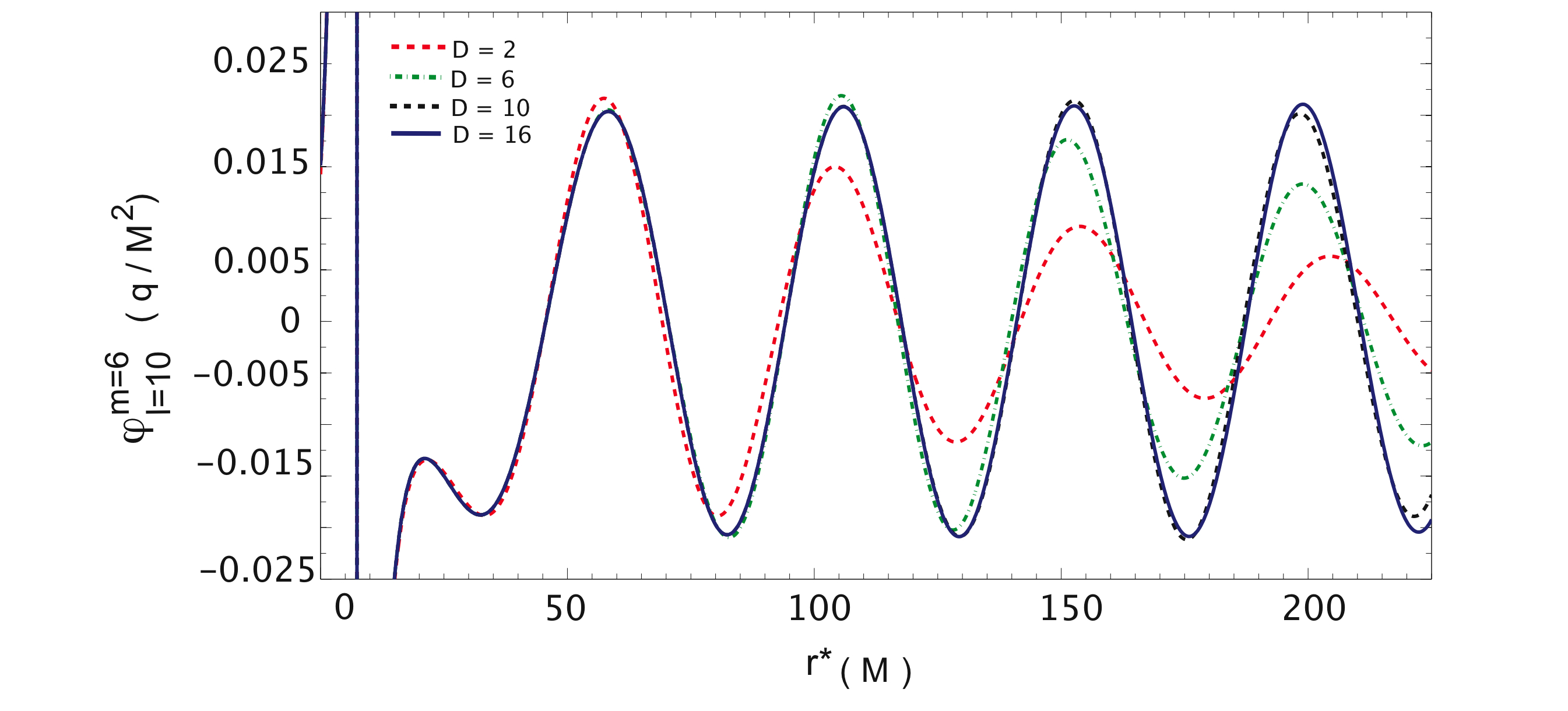}
\caption{We show here three plots of an snapshot of the evolution of the scalar charged particle in 
circular orbital motion at the LSO for the mode $\ell=10$ and $m=6$.  Each plot shows different 
evolutions of the variables $\psi$ (left),
$\phi$ (center), and $\varphi$ (right), for a fixed number of collocation points ($N=50$), but
for different numbers of subdomains ($D=2,6,10,16$).  In this we can see how increasing the
number of subdomains leads to a better accuracy (we can see how the waves get better resolved
and the solution converges as we increase the number of subdomains) with a much less computational
cost as if we would have increase the number of collocation points in a single domain computational
framework.}
\label{multidomainshow}
\end{figure*}

In order to illustrate how the multidomain feature of our numerical framework works, we have performed 
simulations with a fixed number of collocation points but changing the number of subdomains. 
The aim is to show how the solution improves by adding new subdomains.  In our simulations, the
most demanding region for resolution is clearly near the particle.  Apart from this we have the
waves leaving the particle location and going both towards the horizon and towards spatial 
infinity.  These waves are in principle easy to resolve, but for modes with high $\ell$ and
$m$ we have that the source term oscillates like $\exp \{i m \, \Omega_p (t-t_o)\}$ (where
$\Omega_p = \sqrt{M/r^3_p}$ is the coordinate angular velocity of the particle), and hence the  
wavelength gets reduced so that we have many more waves that for low $m$.  These waves are moving
away and need to be resolved.  In Figure~\ref{multidomainshow} we show snapshots of evolutions with 
$50$ collocation points per subdomain, but with different number of subdomains, namely from $2$ to 
$16$ subdomains (half of them to the left of particle and the other half to the right).  The figure
shows the mode $\ell=10$ and $m=6$ (so that the period of the waves is $\pi/(3\Omega_p)$) for
the three fields $(\psi,\phi,\varphi)$.  We can see how for few subdomains the waves are unresolved, 
and that as we increase the number of subdomains the solution converges.  Therefore, the multidomain
structure is a very useful tool to achieve accurate results with a reasonable computational cost.
The key point is to realize what is the optimal number of subdomains, their size, and their 
distribution over the whole computational domain.  Given that in our case the period of the waves
changes with the harmonic number $m$, the optimal strategy for setting the subdomain will depend
on it.  However, since the main aim of this work is to show the principal ingredients of the method 
and to illustrate its performance, we have not explored the possibility of changing the subdomain 
structure as $m$ changes.  However, this is something that should be done for optimizing the 
computational cost in the case of systematic calculations of the self-force.  In this sense, it
would be convenient to have a deep understanding of the computational parameter space in
order to automatically adapt these parameters to the physical parameters and optimize 
in this way the calculations.
This would require an extensive investigation that will be done and presented elsewhere.

The next step in the validation of the code is to compute the components of the self-force
acting on the charged scalar particle moving in circular geodesics around a 
nonrotating black hole.  This is the goal of this numerical scheme, that is, to 
provide accurate computations of the self-force with reasonable computational cost.
The calculation of the self-force require to compute the evolution of the real and imaginary
parts [note that the source term in the evolution equation~(\ref{master}) is complex and hence
both the real and imaginary parts are needed] of the $(\ell,m)$ harmonic components of $\Phi_\ell^m$.
Since $\Phi$ is a real scalar we have that for each $(\ell,m)$ the relation 
$\bar\Phi_\ell^{-m} = (-1)^m\Phi_\ell^m$ holds and hence we do not need to compute the 
modes with $m < 0$.   Given that we need to truncate the sum over $(\ell,m)$ at a given
$\ell_\MAX$, the total number of evolutions that we need to run for a single self-force
calculation is $N_{\mbox{\tiny evolutions}} = [ (\ell_\MAX + 1)(\ell_\MAX + 2)/2]$, where 
here the brackets denote the the nearest integer to the argument.   For the particular
case $\ell_\MAX = 20$, a typical case in our calculations, we need to perform 
$N_{\mbox{\tiny evolutions}} = 231$ evolutions.

In Table~\ref{self_force} we present the results of the computations of the self-force 
vector (actually of the gradient of the regular scalar field) 
acting on a scalar particle in circular (geodesic) motion around a black hole 
at the following radii: $r/M =$ $6$, $7$, $8$, $10$, $14$, and $20$.  For these simulations we have used
our multidomain framework with subdomains that contain $50$ collocation points.  We have
used the same number of subdomains to the right and to the left of the particle.  
The size of these subdomains is typically $\Delta\rsu=20$, specially those near the particle.
As we get far away enough from the particle towards the boundaries (typically at
$\rsu = \pm (500-700)M$), the size of the subdomains is increased since we need less
resolution in those regions.  The total number of
subdomains that we have used ranges from $12-34$.  We have observed that at some point
increasing the number of subdomains (maintaining the size of $\rsu=20M$) does not 
change significantly the results.  Another important point to mention is the fact that
we are computing the self-force right at the particle location.  But since the particle
is at the interface between two subdomains, we obtain two values of the self-force 
components, one from the subdomain to the left [let us call it $\Omega_a$ as in 
equation~(\ref{particleatinterface})],  $\Phi^{\regu,-}_{\alpha} = 
\Phi^{\regu}_{\alpha}(\rsu_{a,R})$, and the other
one from the subdomain to the right ($\Omega_{a+1}$), $\Phi^{\regu,+}_{\alpha} = 
\Phi^{\regu}_{\alpha}(\rsu_{a+1,L})$.  This also provides us with a test of the numerical
calculations as both values have to agree to a good gegree of precision.
In Table~\ref{self_force} we show our results and compare them with two types of 
calculations in the literature: (i) Calculations based on a time-domain method that uses
a characteristic formulation of the scalar field equations~\cite{Haas:2006ne}; (ii) 
calculations based on a frequency-domain method~\cite{DiazRivera:2004ik}.
As we can see, we can get a good numerical approximation to the self-force components 
using a modest amount of computational resources.  It is important to mention that these
computations have a relatively low computational cost (relative to the computational
cost of time-domain simulations of this sort).  The average time for a full self-force
calculation of the type just described in a computer with two Quad-Core Intel Xeon processors 
at $2.8\,$GHz is always in the range $20-30$ minutes.

\begin{table*}
\begin{minipage}{\textwidth}
\caption{\label{self_force} In this table we show the values of the regular field at the particle
location, $(\Phi^{\regu,-}_{\alpha},\Phi^{\regu,+}_{\alpha})$\protect\footnotemark[1], computed at $\varphi_p = 0$.  These results are obtained
for for $\ell_\MAX=20$, $N=50$, and $D = 12-36$.  The minimum size of the subdomains, as measured 
in terms of the tortoise coordinate is $\Delta\rsu= 20 M$, which corresponds to the subdomains near 
the particle location.  The table shows our numerical results for different circular orbits 
($r/M =$ $6$, $7$, $8$, $10$, $14$, and $20$) and the relative errors with respect to the results obtained with a time-domain 
method in~\cite{Haas:2007kz}, and with a frequency-domain method in~\cite{DiazRivera:2004ik} and~\cite{Haas:2006ne}.} 
\centering    
\begin{ruledtabular}
\begin{tabular}{c|c|c|c|c|c}  
\multirow{3}{*}{$r (M)$} & Component & Estimation using & Estimation from  & Error relative to & Error relative to \\
          & of $\Phi^{\regu}_\alpha$ & the PSC Method   & Frequency-domain & Frequency-domain   & Time-domain   \\
          &                         &                   & calculations in~\cite{DiazRivera:2004ik} and~\cite{Haas:2006ne} & calculations in~\cite{DiazRivera:2004ik} and~\cite{Haas:2006ne} & calculations in~\cite{Haas:2007kz} \\ 
\hline
\multirow{3}{*}{6} & $(\Phi^{\regu,-}_t,\Phi^{\regu,+}_t)$ & $(3.60777, 3.60778)\cdot 10^{-4}$  & $3.609072\cdot 10^{-4}$ & $(0.03,0.03)$ \%  &  $(0.12,0.12)$ \% \\
                   & $(\Phi^{\regu,-}_r,\Phi^{\regu,+}_r)$ & $(1.67364, 1.67362)\cdot 10^{-4}$  & $1.67728\cdot 10^{-4}$ & $(0.2,0.2)$ \%    &  $(0.18,0.18)$ \%   \\
                   & $(\Phi^{\regu,-}_\varphi,\Phi^{\regu,+}_\varphi)$ & $(-5.30422, -5.30438)\cdot 10^{-3}$ & $-5.304231\cdot 10^{-3}$ & $(4\cdot 10^{-4},  10^{-3})$ \%  &  $(6\cdot 10^{-4}, 10^{-3})$ \% \\
\hline
\multirow{3}{*}{7} & $(\Phi^{\regu,-}_t,\Phi^{\regu,+}_t)$ & $(1.76638, 1.76639)\cdot 10^{-4}$  &                        &                   &                   \\
                   & $(\Phi^{\regu,-}_r,\Phi^{\regu,+}_r)$ & $(7.84007, 7.84001)\cdot 10^{-5}$  & $7.85067\cdot 10^{-5}$ & $(0.13,0.13)$ \%  &                   \\
                   & $(\Phi^{\regu,-}_\varphi,\Phi^{\regu,+}_\varphi)$ & $(-3.2730, -3.2733)\cdot 10^{-3}$ &             &                   &                    \\
\hline
\multirow{3}{*}{8} & $(\Phi^{\regu,-}_t,\Phi^{\regu,+}_t)$ & $(9.76454, 9.76457)\cdot 10^{-5}$  &                        &                   &                    \\
                   & $(\Phi^{\regu,-}_r,\Phi^{\regu,+}_r)$ & $(4.0835, 4.0832)\cdot 10^{-5}$  & $4.08250\cdot 10^{-5}$   & $(0.02,0.01)$ \%  &                    \\
                   & $(\Phi^{\regu,-}_\varphi,\Phi^{\regu,+}_\varphi)$ & $(-2.2115, -2.2108)\cdot 10^{-3}$ &             &                   &                   \\
\hline
\multirow{3}{*}{10} & $(\Phi^{\regu,-}_t,\Phi^{\regu,+}_t)$ & $(3.74362, 3.74363)\cdot 10^{-5}$ &                        &                   &                   \\
                   & $(\Phi^{\regu,-}_r,\Phi^{\regu,+}_r)$ & $(1.3804, 1.3801)\cdot 10^{-5}$    & $1.37844\cdot 10^{-5}$ & $(0.14,0.12)$ \%  &                   \\
                   & $(\Phi^{\regu,-}_\varphi,\Phi^{\regu,+}_\varphi)$ & $(-1.1860, -1.1858)\cdot 10^{-3}$    &          &                   &                   \\
\hline
\multirow{3}{*}{14} & $(\Phi^{\regu,-}_t,\Phi^{\regu,+}_t)$ & $(9.176912, 9.176914)\cdot 10^{-6}$  &                       &                   &                   \\
                   & $(\Phi^{\regu,-}_r,\Phi^{\regu,+}_r)$ & $(2.7269, 2.7266)\cdot 10^{-6}$  & $2.72008\cdot 10^{-6}$ & $(0.25,0.24)$ \%    &                     \\
                   & $(\Phi^{\regu,-}_\varphi,\Phi^{\regu,+}_\varphi)$ & $(-4.8383, -4.8389)\cdot 10^{-4}$    &        &                   &                   \\
\hline
\multirow{3}{*}{20} & $(\Phi^{\regu,-}_t,\Phi^{\regu,+}_t)$ & $(2.08859, 2.08858)\cdot 10^{-6}$  &                       &                   &                   \\
                   & $(\Phi^{\regu,-}_r,\Phi^{\regu,+}_r)$ & $(4.9444, 4.9440)\cdot 10^{-7}$  & $4.93790\cdot 10^{-7}$ & $(0.13,0.12)$ \%    &                  \\
                   & $(\Phi^{\regu,-}_\varphi,\Phi^{\regu,+}_\varphi)$ & $(-1.92449, -1.92436)\cdot 10^{-4}$    &        &                   &                   \\
\end{tabular} 
\end{ruledtabular}
\footnotetext[1]{We show the values of the
gradient of the regular field instead of the components of the self-force for the sake of comparing
with other results in the literature.}
\end{minipage}
\end{table*} 

\section{Conclusions and Discussion}
\label{discussion}

In this paper we have introduced a new time-domain technique for
the simulations of EMRIs.  The main ingredient of the method is to use a
multi-domain framework in which the SCO, described as a point-like object,
is located at the interface between two subdomains.  In this way we 
have shown that this technique enjoys the exponential convergence 
property of the PSC method and its accuracy and, at the same time,
it is also an efficient method to make time-domain computations of
the self-force.  We have shown that we can achieve a good accuracy 
in this computations by using a relatively low number of collocation points. 
In this sense, the multidomain framework allows us to locate more
collocation points in the region where they are most needed, i.e. 
around the particle, by choosing appropriately the number of subdomains,
and their size and location.  Another positive property of our
numerical scheme is that it can be easily parallelized to be used
in supercomputers, as we can assign the work of each subdomain to
different CPUs.  The only information that we need to communicate
is the one necessary to satisfy the matching conditions between
subdomains.

The calculations performed for this paper show the potential of this 
technique for the description of EMRIs.  These results still leave
room for improvement as we have not explore yet the full parameter
space of the numerical method (number of collocation points per subdomain,
number and distribution of subdomains, parameters of the spectral filter, 
penalty parameters, etc), which is wide enough.  In this regard, it
would be convenient to be able to adjust the computational parameters
to the physical parameters of each mode in an automatic way. 
In addition to this, there
are several additions/modifications to the present numerical framework
that may improve the present accuracy and
efficiency of the calculations, some of which we will investigated
in future calculations by the present authors.  We list here some 
of them: (i) {\em Compactification of the computational
domain}.  We can change the mapping between the spectral and physical
domains incorporating the two infinities (the horizon location at
$\rsu_\Hor\rightarrow-\infty$ and spatial infinity at 
$\rsu_\Inf\rightarrow\infty$) into the calculation.  In practice,
this mapping only would need to be performed in the first and last
subdomains.  By doing this, the outgoing boundary conditions (\ref{outgoingbcs}) 
would be exact instead of approximate as they are know.  An alternative to this 
could be to improve the boundary conditions by analyzing the solution near 
the two infinities, like for instance in~\cite{Barack:2008ms} in a frequency 
domain framework.  (ii) {\em Reduction of the time step}.  As we have mentioned
before, the PSC method has a more strict CFL condition as a FD scheme, namely
$\Delta t \sim N^{-2}$ versus $\Delta t \sim N^{-1}$.  This is because of
high density of collocation points near the boundaries and the fact that
the minimum time step come from the minimum separation between points.  
A way of changing this is again to change the mapping  between the spectral
and physical domains.  One known way of doing this is the modification
introduced by Kosloff and Tal-Ezer~\cite{Kosloff:1993kt}, which was shown to 
lead to a time step restriction like in the FD case, that is $\Delta t \sim N^{-1}$.  
(iii) {\em Richardson extrapolation}.  In our calculations we have computed
harmonic components of the scalar field, $\Phi_\ell^m$, up to a certain
$\ell^{}_{max}$, in this work $\ell^{}_{max} = 15-20$.  In other words, we
have truncated the expansion in spherical harmonics.  We can try to improve
the final numerical results for the scalar field and derived quantities,
like the self-force, by profiting from our analytical knowledge of the expansions 
in inverse powers of the harmonic number $\ell$.  This can be done using
numerical techniques based on the Richardson extrapolation, like 
in~\cite{Detweiler:2002gi}, which can provide an estimation of the different
quantities of interest as $\ell^{}_{max}\rightarrow\infty$.

Beyond the case studied in this paper, circular orbits of a charged scalar
particle, we will study in the future how to extend our time-domain
computational framework in order to include generic (eccentric) orbits.
The main difficulty is obviously to deal with a particle moving the radial
direction as our present framework assumes that it is located at a fixed
value (circular motion).  There are several ways in which we can attack 
this problem.  One is to try to implement some type of moving grid 
technique (as it was done in a similar situation in~\cite{Sopuerta:2005gz}),
but the reconstruction of the grid at every time step could increase 
significantly the computational cost of the evolution.  Another possibility
is to try to make a change of coordinates to coordinates comoving with the
particle, so that the techniques presented in this paper can be implemented
in a straightforward way.

On the other hand, we expect that MBHs sitting at galactic centers will have
considerable spins ($a/M \sim 0.7$ or bigger, where $a$ is the Kerr spin
parameter) and hence the MBH should be described by the Kerr metric instead
of the Schwarzschild metric.  This means to have a less symmetric background, 
instead of the spherical symmetry of Schwarzschild just the axisymmetry of
Kerr.  In that case one can separate the dependence on the azimuthal angle and
is left with $2+1$ wave-type equations with singular terms.  The main difficulty
in this case is that each $m$-mode (coming from the separation of the azimuthal
angle) diverges logarithmically at the particle location.  Then, before trying
to transfer the techniques presented in this paper to the Kerr case, one has
to apply before a regularization procedure as it has been done in~\cite{Barack:2007jh,Vega:2007mc,Lousto:2008mb}.
Apart from this, we expect that most of the methods presented in this paper
will be helpful in achieving efficient simulations of EMRIs in the case of
a spinning  MBH.

\acknowledgments 
We would like to thank Leor Barack, Jos\'e Luis Jaramillo and Eric Poisson 
for helpful discussions and encouragement. 
CFS acknowledges support from the Ram\'on y Cajal Programme of the
Ministry of Education and Science of Spain and by a Marie Curie
International Reintegration Grant (MIRG-CT-2007-205005/PHY) within the
7th European Community Framework Programme. 
PCM is supported by a predoctoral FPU fellowship of the Spanish Ministry of
Science and Innovation (MICINN).
Financial support from the Spanish Ministry of Science and Education 
contract ESP2007-61712 is gratefully acknowledged.
This research has been partly performed using the resources of the Centre de 
Supercomputaci\'o de Catalunya (CESCA).

\appendix

\section{Special Functions used in this work}
\label{specialfunctions}
Here, we summarize the main conventions used for the special functions involved
in calculations of this paper.

\subsection{Spherical Harmonics}
\label{sphericalharmonics}
The expression we use for the scalar spherical harmonics is:
\begin{equation}
\label{ylm}
Y_{\ell}^{m}(\theta,\varphi) = \sqrt{\frac{2 \ell + 1}{4 \pi} 
      \frac{\left(\ell - m\right)!}{\left(\ell + m\right)!}}\,
      P^m_\ell(\cos{\theta}) e^{i m \varphi}\,,
\end{equation}
where $P^m_\ell$ are the associated Legendre polynomials [we use the same 
expressions as in~\cite{Abramowitz:1970as}, equations (8.6.6) and 
(8.6.18)]
\begin{equation}
P^m_\ell(x) = \frac{(-1)^{\ell+m}}{2^\ell\,\ell!}(1-x^2)^{m/2}\frac{d^{\ell+m}}{dx^{\ell+m}}
(1-x^2)^\ell \,,
\end{equation}
where $\ell$ is a non-negative integer and $m$ is an integer restricted to 
the following range: $m \in (-\ell\,,-\ell+1\,,\ldots\,,\ell-1\,,\ell)\,$.

\subsection{Chebyshev Polynomials}
\label{chebyshev}
The Chebyshev polynomials can be expresses as follows:
\begin{eqnarray}
T^{}_n(X) =\cos\left(n\cos^{-1}(X) \right),
\end{eqnarray}
and are defined in the interval $\left[-1,1\right]$ with 
$\vert T^{}_n(X)\vert\leq 1$, where $n$ is the degree of 
the polynomial.  Chebyshev polynomials are orthogonal in
the {\em continuum} in the following sense:
\begin{equation}
\left(T^{}_n,T^{}_m\right) = \int^{1}_{-1} \frac{dX}{\sqrt{1-X^2}}
T^{}_n(X)\,T^{}_m(X) = \frac{\pi c^{}_{n}}{2}\delta^{}_{nm}\,,
\end{equation}
where the coefficients $c^{}_{n}$ are given in equation~(\ref{cns}).

The set of collocation points for the spatial discretization of our
PDEs by means of the PSC method are the extrema of the Chebyshev 
polynomials, together with the end points $X=\pm1$ [that is, the zeros
of the polynomial given in equation~\ref{chebyshevlobattogrid})].  
These points form, in the interval $[-1,1]$, a Chebyshev-Lobatto 
collocation grid.  The cardinal functions associated with them 
are~$(i=0,\ldots,N)$:
\begin{equation}
{\cal C}^{}_i(X) = \frac{(1-X^2){T'_N}(X)}
{(1-X^{2}_i)(X-X^{}_i){T''_N}(X^{}_i)}\,.
\end{equation}
Once this set of collocation points is adopted, and taking into
account the properties of the Gauss-Lobatto-Chebyshev quadratures
(see, e.g.~\cite{Boyd}), the Chebyshev polynomials have another
orthogonality relation, this time in the {\em discrete}, in the
following sense ($n,m = 0,\ldots,N$):
\begin{eqnarray}
\left[ T^{}_n, T^{}_m\right]=  \sum_{i=0}^{N} w^{}_i\, T^{}_n(X^{}_i)\, 
T^{}_m(X^{}_{i}) = 
\nu_n^2\,\delta^{}_{nm}\,,  
\label{scalarp}
\end{eqnarray}
where $w^{}_i$ are the weights associated with the Chebyshev-Lobatto 
grid, $w^{}_i = \pi/(N\,\bar{c}^{}_{i})$, and where the $\bar{c}^{}_i$'s 
are normalization coefficients given by
\begin{eqnarray}
\bar{c}^{}_i=\left\{\begin{array}{ll} 2  & \text{for}~i=0,N\,, \\[2mm]
                                      1  & \text{otherwise\,.}
\end{array}\right.
\end{eqnarray}
Finally, the constants $\nu^{}_n$ in (\ref{scalarp}) are given by
$\nu^{2}_n=\pi \bar{c}^{}_n/2$.

On the other hand, introducing a new variable, $X =\cos\theta$, the Chebyshev polynomials
look like
\begin{equation}
T^{}_{n}(\cos\theta) = \cos(n\theta)\,. \label{changespectralcoord}
\end{equation}
Then, an expansion in Chebyshev polynomials can be mapped to a cosine
expansion.

\section{Spectral filter}
\label{exponentialfilter}
In order to reduce the spurious high-frequency components of our 
numerical solutions, we apply a spectral filter of the exponential
type to the solution after every time step, that is, after every
full RK step.  The scheme for the action of the spectral filter is
\begin{eqnarray}
\{\mb{U}^{}_i\} \;\stackrel{FFT}{\longrightarrow}\; 
\{\mb{a}^{}_n\} \;\stackrel{\text{Filter}}{\longrightarrow}\;
\{\tilde{\mb{a}}{}^{}_{n}\} \;\stackrel{FFT}{\longrightarrow}
\{{\tilde{\mb{U}}}{}^{}_{i}\}\,,
\end{eqnarray}
where $\{\mb{U}^{}_i\}$ are the values of the solutions at the
collocation points after the RK step; $\{\mb{a}^{}_n\}$ are their
corresponding spectral components; $\{\mb{b}^{}_n\}$ are the
filtered spectral components; and $\{{\tilde{\mb{U}}}{}^{}_{i}\}$
are the filtered values of the solution at the collocation points.
The exponential filter is defined by its action on the spectral
coefficients $\{\mb{a}^{}_n\}$ to yield the spectral coefficients
$\{\mb{b}^{}_n\}$.  This action is given by
\begin{eqnarray}
{\tilde {\mb{a}}}^{}_n = \sigma\left(\frac{n}{N}\right)\,
\mb{a}^{}_n \,,
\end{eqnarray}
where $\sigma(n/N)$ is the exponential filter 
\begin{eqnarray}
\sigma\left(\frac{n}{N}\right) = \left\{
\begin{array}{cl}  1  & \mbox{for~} 0\leqslant n\leqslant N^{}_c\,, \\ 
 \exp\left[ -\alpha ( \frac{n-N^{}_c}{N-N^{}_c} )^{\gamma}\right] & 
 \mbox{for~} N^{}_c< n\leq N\,, \end{array} \right. 
\end{eqnarray}
where $N^{}_c$ is the cut-off mode number, $\gamma$ is the order of the filter 
(typically chosen to be of the order of the number of collocation
points, $N$), and $\alpha$ is the machine accuracy parameter, which is related 
to the machine accuracy, $\epsilon^{}_M$, by $\alpha=-\ln \epsilon^{}_M$.


\begin{thebibliography}{76}
\expandafter\ifx\csname natexlab\endcsname\relax\def\natexlab#1{#1}\fi
\expandafter\ifx\csname bibnamefont\endcsname\relax
  \def\bibnamefont#1{#1}\fi
\expandafter\ifx\csname bibfnamefont\endcsname\relax
  \def\bibfnamefont#1{#1}\fi
\expandafter\ifx\csname citenamefont\endcsname\relax
  \def\citenamefont#1{#1}\fi
\expandafter\ifx\csname url\endcsname\relax
  \def\url#1{\texttt{#1}}\fi
\expandafter\ifx\csname urlprefix\endcsname\relax\def\urlprefix{URL }\fi
\providecommand{\bibinfo}[2]{#2}
\providecommand{\eprint}[2][]{\url{#2}}

\bibitem[{LIS()}]{LISA}
\emph{\bibinfo{title}{{LISA}}}, \bibinfo{note}{{\tt http://www.esa.int/science/lisa}, {\tt
  http://lisa.jpl.nasa.gov}}.

\bibitem[{\citenamefont{Gair et~al.}(2004)\citenamefont{Gair, Barack,
  Creighton, Cutler, Larson, Phinney, and Vallisneri}}]{Gair:2004iv}
\bibinfo{author}{\bibfnamefont{J.~R.} \bibnamefont{Gair}},
  \bibinfo{author}{\bibfnamefont{L.}~\bibnamefont{Barack}},
  \bibinfo{author}{\bibfnamefont{T.}~\bibnamefont{Creighton}},
  \bibinfo{author}{\bibfnamefont{C.}~\bibnamefont{Cutler}},
  \bibinfo{author}{\bibfnamefont{S.~L.} \bibnamefont{Larson}},
  \bibinfo{author}{\bibfnamefont{E.~S.} \bibnamefont{Phinney}},
  \bibnamefont{and}
  \bibinfo{author}{\bibfnamefont{M.}~\bibnamefont{Vallisneri}},
  \bibinfo{journal}{Class. Quant. Grav.} \textbf{\bibinfo{volume}{21}},
  \bibinfo{pages}{S1595} (\bibinfo{year}{2004}), \eprint{gr-qc/0405137}.

\bibitem[{\citenamefont{Hopman and Alexander}(2006)}]{Hopman:2006xn}
\bibinfo{author}{\bibfnamefont{C.}~\bibnamefont{Hopman}} \bibnamefont{and}
  \bibinfo{author}{\bibfnamefont{T.}~\bibnamefont{Alexander}},
  \bibinfo{journal}{Astrophys. J.} \textbf{\bibinfo{volume}{645}},
  \bibinfo{pages}{L133} (\bibinfo{year}{2006}), \eprint{astro-ph/0603324}.

\bibitem[{\citenamefont{Sigurdsson and Rees}(1996)}]{Sigurdsson:1996uz}
\bibinfo{author}{\bibfnamefont{S.}~\bibnamefont{Sigurdsson}} \bibnamefont{and}
  \bibinfo{author}{\bibfnamefont{M.~J.} \bibnamefont{Rees}}
  (\bibinfo{year}{1996}), \eprint{astro-ph/9608093}.

\bibitem[{\citenamefont{Amaro-Seoane et~al.}(2007)}]{AmaroSeoane:2007aw}
\bibinfo{author}{\bibfnamefont{P.}~\bibnamefont{Amaro-Seoane}}
  \bibnamefont{et~al.}, \bibinfo{journal}{Class. Quant. Grav.}
  \textbf{\bibinfo{volume}{24}}, \bibinfo{pages}{R113} (\bibinfo{year}{2007}),
  \eprint{astro-ph/0703495}.

\bibitem[{\citenamefont{Brown et~al.}(2007)}]{Brown06}
\bibinfo{author}{\bibfnamefont{D.~A.} \bibnamefont{Brown}}
  \bibnamefont{et~al.}, \bibinfo{journal}{Phys. Rev. Lett.}
  \textbf{\bibinfo{volume}{99}}, \bibinfo{pages}{201102}
  (\bibinfo{year}{2007}), \eprint{gr-qc/0612060}.

\bibitem[{\citenamefont{Mandel et~al.}(2007)\citenamefont{Mandel, Brown, Gair,
  and Miller}}]{Mandel:2007hi}
\bibinfo{author}{\bibfnamefont{I.}~\bibnamefont{Mandel}},
  \bibinfo{author}{\bibfnamefont{D.~A.} \bibnamefont{Brown}},
  \bibinfo{author}{\bibfnamefont{J.~R.} \bibnamefont{Gair}}, \bibnamefont{and}
  \bibinfo{author}{\bibfnamefont{M.~C.} \bibnamefont{Miller}}
  (\bibinfo{year}{2007}), \eprint{0705.0285}.

\bibitem[{Adv({\natexlab{a}})}]{AdvLIGO}
\emph{\bibinfo{title}{{Advanced LIGO}}}, \bibinfo{note}{{\tt
  http://www.ligo.caltech.edu/advLIGO}}.

\bibitem[{Adv({\natexlab{b}})}]{AdvVIRGO}
\emph{\bibinfo{title}{{Advanced VIRGO}}}, \bibinfo{note}{{\tt
  http://wwwcascina.virgo.infn.it/advirgo}}.

\bibitem[{\citenamefont{Finn and Thorne}(2000)}]{Finn:2000sy}
\bibinfo{author}{\bibfnamefont{L.~S.} \bibnamefont{Finn}} \bibnamefont{and}
  \bibinfo{author}{\bibfnamefont{K.~S.} \bibnamefont{Thorne}},
  \bibinfo{journal}{Phys. Rev.} \textbf{\bibinfo{volume}{D62}},
  \bibinfo{pages}{124021} (\bibinfo{year}{2000}), \eprint{gr-qc/0007074}.

\bibitem[{\citenamefont{Poisson}(2004)}]{Poisson:2004lr}
\bibinfo{author}{\bibfnamefont{E.}~\bibnamefont{Poisson}},
  \bibinfo{journal}{Living Rev. Relativity} \textbf{\bibinfo{volume}{7}},
  \bibinfo{pages}{6} (\bibinfo{year}{2004}), \eprint{gr-qc/0306052},
  \urlprefix\url{http://www.livingreviews.org/lrr-2004-6}.

\bibitem[{\citenamefont{Tanaka}(2006)}]{Tanaka:2005ue}
\bibinfo{author}{\bibfnamefont{T.}~\bibnamefont{Tanaka}},
  \bibinfo{journal}{Prog. Theor. Phys. Suppl.} \textbf{\bibinfo{volume}{163}},
  \bibinfo{pages}{120} (\bibinfo{year}{2006}), \eprint{gr-qc/0508114}.

\bibitem[{\citenamefont{Glampedakis}(2005)}]{Glampedakis:2005hs}
\bibinfo{author}{\bibfnamefont{K.}~\bibnamefont{Glampedakis}},
  \bibinfo{journal}{Class. Quant. Grav.} \textbf{\bibinfo{volume}{22}},
  \bibinfo{pages}{S605} (\bibinfo{year}{2005}), \eprint{gr-qc/0509024}.

\bibitem[{\citenamefont{Hughes et~al.}(2005)\citenamefont{Hughes, Drasco,
  Flanagan, and Franklin}}]{Hughes:2005qb}
\bibinfo{author}{\bibfnamefont{S.~A.} \bibnamefont{Hughes}},
  \bibinfo{author}{\bibfnamefont{S.}~\bibnamefont{Drasco}},
  \bibinfo{author}{\bibfnamefont{E.~E.} \bibnamefont{Flanagan}},
  \bibnamefont{and} \bibinfo{author}{\bibfnamefont{J.}~\bibnamefont{Franklin}},
  \bibinfo{journal}{Phys. Rev. Lett.} \textbf{\bibinfo{volume}{94}},
  \bibinfo{pages}{221101} (\bibinfo{year}{2005}), \eprint{gr-qc/0504015}.

\bibitem[{\citenamefont{Drasco and Hughes}(2006)}]{Drasco:2005kz}
\bibinfo{author}{\bibfnamefont{S.}~\bibnamefont{Drasco}} \bibnamefont{and}
  \bibinfo{author}{\bibfnamefont{S.~A.} \bibnamefont{Hughes}},
  \bibinfo{journal}{Phys. Rev.} \textbf{\bibinfo{volume}{D73}},
  \bibinfo{pages}{024027} (\bibinfo{year}{2006}), \eprint{gr-qc/0509101}.

\bibitem[{\citenamefont{Sago et~al.}(2005)\citenamefont{Sago, Tanaka, Hikida,
  and Nakano}}]{Sago:2005gd}
\bibinfo{author}{\bibfnamefont{N.}~\bibnamefont{Sago}},
  \bibinfo{author}{\bibfnamefont{T.}~\bibnamefont{Tanaka}},
  \bibinfo{author}{\bibfnamefont{W.}~\bibnamefont{Hikida}}, \bibnamefont{and}
  \bibinfo{author}{\bibfnamefont{H.}~\bibnamefont{Nakano}},
  \bibinfo{journal}{Prog. Theor. Phys.} \textbf{\bibinfo{volume}{114}},
  \bibinfo{pages}{509} (\bibinfo{year}{2005}), \eprint{gr-qc/0506092}.

\bibitem[{\citenamefont{Ganz et~al.}(2007)\citenamefont{Ganz, Hikida, Nakano,
  Sago, and Tanaka}}]{Ganz:2007rf}
\bibinfo{author}{\bibfnamefont{K.}~\bibnamefont{Ganz}},
  \bibinfo{author}{\bibfnamefont{W.}~\bibnamefont{Hikida}},
  \bibinfo{author}{\bibfnamefont{H.}~\bibnamefont{Nakano}},
  \bibinfo{author}{\bibfnamefont{N.}~\bibnamefont{Sago}}, \bibnamefont{and}
  \bibinfo{author}{\bibfnamefont{T.}~\bibnamefont{Tanaka}}
  (\bibinfo{year}{2007}), \eprint{gr-qc/0702054}.

\bibitem[{\citenamefont{Mino}(2003)}]{Mino:2003yg}
\bibinfo{author}{\bibfnamefont{Y.}~\bibnamefont{Mino}}, \bibinfo{journal}{Phys.
  Rev.} \textbf{\bibinfo{volume}{D67}}, \bibinfo{pages}{084027}
  (\bibinfo{year}{2003}), \eprint{gr-qc/0302075}.

\bibitem[{\citenamefont{Pound et~al.}(2005)\citenamefont{Pound, Poisson, and
  Nickel}}]{Pound:2005fs}
\bibinfo{author}{\bibfnamefont{A.}~\bibnamefont{Pound}},
  \bibinfo{author}{\bibfnamefont{E.}~\bibnamefont{Poisson}}, \bibnamefont{and}
  \bibinfo{author}{\bibfnamefont{B.~G.} \bibnamefont{Nickel}},
  \bibinfo{journal}{Phys. Rev.} \textbf{\bibinfo{volume}{D72}},
  \bibinfo{pages}{124001} (\bibinfo{year}{2005}), \eprint{gr-qc/0509122}.

\bibitem[{\citenamefont{Pound and Poisson}(2008)}]{Pound:2007ti}
\bibinfo{author}{\bibfnamefont{A.}~\bibnamefont{Pound}} \bibnamefont{and}
  \bibinfo{author}{\bibfnamefont{E.}~\bibnamefont{Poisson}},
  \bibinfo{journal}{Phys. Rev.} \textbf{\bibinfo{volume}{D77}},
  \bibinfo{pages}{044012} (\bibinfo{year}{2008}), \eprint{0708.3037}.

\bibitem[{\citenamefont{Mino}(2008)}]{Mino:2007ft}
\bibinfo{author}{\bibfnamefont{Y.}~\bibnamefont{Mino}}, \bibinfo{journal}{Phys.
  Rev.} \textbf{\bibinfo{volume}{D77}}, \bibinfo{pages}{044008}
  (\bibinfo{year}{2008}), \eprint{0711.3007}.

\bibitem[{\citenamefont{Hinderer and Flanagan}(2008)}]{Hinderer:2008dm}
\bibinfo{author}{\bibfnamefont{T.}~\bibnamefont{Hinderer}} \bibnamefont{and}
  \bibinfo{author}{\bibfnamefont{E.~E.} \bibnamefont{Flanagan}}
  (\bibinfo{year}{2008}), \eprint{0805.3337}.

\bibitem[{\citenamefont{Mino et~al.}(1997)\citenamefont{Mino, Sasaki, and
  Tanaka}}]{Mino:1997nk}
\bibinfo{author}{\bibfnamefont{Y.}~\bibnamefont{Mino}},
  \bibinfo{author}{\bibfnamefont{M.}~\bibnamefont{Sasaki}}, \bibnamefont{and}
  \bibinfo{author}{\bibfnamefont{T.}~\bibnamefont{Tanaka}},
  \bibinfo{journal}{Phys. Rev.} \textbf{\bibinfo{volume}{D55}},
  \bibinfo{pages}{3457} (\bibinfo{year}{1997}), \eprint{gr-qc/9606018}.

\bibitem[{\citenamefont{Quinn and Wald}(1997)}]{Quinn:1997am}
\bibinfo{author}{\bibfnamefont{T.~C.} \bibnamefont{Quinn}} \bibnamefont{and}
  \bibinfo{author}{\bibfnamefont{R.~M.} \bibnamefont{Wald}},
  \bibinfo{journal}{Phys. Rev.} \textbf{\bibinfo{volume}{D56}},
  \bibinfo{pages}{3381} (\bibinfo{year}{1997}), \eprint{gr-qc/9610053}.

\bibitem[{\citenamefont{Detweiler and Whiting}(2003)}]{Detweiler:2002mi}
\bibinfo{author}{\bibfnamefont{S.}~\bibnamefont{Detweiler}} \bibnamefont{and}
  \bibinfo{author}{\bibfnamefont{B.~F.} \bibnamefont{Whiting}},
  \bibinfo{journal}{Phys. Rev.} \textbf{\bibinfo{volume}{D67}},
  \bibinfo{pages}{024025} (\bibinfo{year}{2003}), \eprint{gr-qc/0202086}.

\bibitem[{\citenamefont{Gralla and Wald}(2008)}]{Gralla:2008fg}
\bibinfo{author}{\bibfnamefont{S.~E.} \bibnamefont{Gralla}} \bibnamefont{and}
  \bibinfo{author}{\bibfnamefont{R.~M.} \bibnamefont{Wald}},
  \bibinfo{journal}{Class. Quant. Grav.} \textbf{\bibinfo{volume}{25}},
  \bibinfo{pages}{205009} (\bibinfo{year}{2008}), \eprint{0806.3293}.

\bibitem[{\citenamefont{Barut}(1980)}]{Barut:1980ao}
\bibinfo{author}{\bibfnamefont{A.~O.} \bibnamefont{Barut}},
  \emph{\bibinfo{title}{Electrodynamics and Classical Theory of Fields and
  Particles}} (\bibinfo{publisher}{Dover}, \bibinfo{address}{New York},
  \bibinfo{year}{1980}).

\bibitem[{\citenamefont{Jackson}(1999)}]{Jackson:99jd}
\bibinfo{author}{\bibfnamefont{J.~D.} \bibnamefont{Jackson}},
  \emph{\bibinfo{title}{Classical Electrodynamics}} (\bibinfo{publisher}{John
  Wiley \& Sons}, \bibinfo{address}{New York}, \bibinfo{year}{1999}),
  \bibinfo{edition}{3rd} ed.

\bibitem[{\citenamefont{Barack and Ori}(2000)}]{Barack:1999wf}
\bibinfo{author}{\bibfnamefont{L.}~\bibnamefont{Barack}} \bibnamefont{and}
  \bibinfo{author}{\bibfnamefont{A.}~\bibnamefont{Ori}},
  \bibinfo{journal}{Phys. Rev.} \textbf{\bibinfo{volume}{D61}},
  \bibinfo{pages}{061502} (\bibinfo{year}{2000}), \eprint{gr-qc/9912010}.

\bibitem[{\citenamefont{Barack}(2000)}]{Barack:2000eh}
\bibinfo{author}{\bibfnamefont{L.}~\bibnamefont{Barack}},
  \bibinfo{journal}{Phys. Rev.} \textbf{\bibinfo{volume}{D62}},
  \bibinfo{pages}{084027} (\bibinfo{year}{2000}), \eprint{gr-qc/0005042}.

\bibitem[{\citenamefont{Barack}(2001)}]{Barack:2001bw}
\bibinfo{author}{\bibfnamefont{L.}~\bibnamefont{Barack}},
  \bibinfo{journal}{Phys. Rev.} \textbf{\bibinfo{volume}{D64}},
  \bibinfo{pages}{084021} (\bibinfo{year}{2001}), \eprint{gr-qc/0105040}.

\bibitem[{\citenamefont{Mino et~al.}(2003)\citenamefont{Mino, Nakano, and
  Sasaki}}]{Mino:2001mq}
\bibinfo{author}{\bibfnamefont{Y.}~\bibnamefont{Mino}},
  \bibinfo{author}{\bibfnamefont{H.}~\bibnamefont{Nakano}}, \bibnamefont{and}
  \bibinfo{author}{\bibfnamefont{M.}~\bibnamefont{Sasaki}},
  \bibinfo{journal}{Prog. Theor. Phys.} \textbf{\bibinfo{volume}{108}},
  \bibinfo{pages}{1039} (\bibinfo{year}{2003}), \eprint{gr-qc/0111074}.

\bibitem[{\citenamefont{Barack et~al.}(2002)\citenamefont{Barack, Mino, Nakano,
  Ori, and Sasaki}}]{Barack:2001gx}
\bibinfo{author}{\bibfnamefont{L.}~\bibnamefont{Barack}},
  \bibinfo{author}{\bibfnamefont{Y.}~\bibnamefont{Mino}},
  \bibinfo{author}{\bibfnamefont{H.}~\bibnamefont{Nakano}},
  \bibinfo{author}{\bibfnamefont{A.}~\bibnamefont{Ori}}, \bibnamefont{and}
  \bibinfo{author}{\bibfnamefont{M.}~\bibnamefont{Sasaki}},
  \bibinfo{journal}{Phys. Rev. Lett.} \textbf{\bibinfo{volume}{88}},
  \bibinfo{pages}{091101} (\bibinfo{year}{2002}), \eprint{gr-qc/0111001}.

\bibitem[{\citenamefont{Barack and Ori}(2002)}]{Barack:2002mha}
\bibinfo{author}{\bibfnamefont{L.}~\bibnamefont{Barack}} \bibnamefont{and}
  \bibinfo{author}{\bibfnamefont{A.}~\bibnamefont{Ori}},
  \bibinfo{journal}{Phys. Rev.} \textbf{\bibinfo{volume}{D66}},
  \bibinfo{pages}{084022} (\bibinfo{year}{2002}), \eprint{gr-qc/0204093}.

\bibitem[{\citenamefont{Detweiler et~al.}(2003)\citenamefont{Detweiler,
  Messaritaki, and Whiting}}]{Detweiler:2002gi}
\bibinfo{author}{\bibfnamefont{S.}~\bibnamefont{Detweiler}},
  \bibinfo{author}{\bibfnamefont{E.}~\bibnamefont{Messaritaki}},
  \bibnamefont{and} \bibinfo{author}{\bibfnamefont{B.~F.}
  \bibnamefont{Whiting}}, \bibinfo{journal}{Phys. Rev.}
  \textbf{\bibinfo{volume}{D67}}, \bibinfo{pages}{104016}
  (\bibinfo{year}{2003}), \eprint{gr-qc/0205079}.

\bibitem[{\citenamefont{Haas and Poisson}(2006)}]{Haas:2006ne}
\bibinfo{author}{\bibfnamefont{R.}~\bibnamefont{Haas}} \bibnamefont{and}
  \bibinfo{author}{\bibfnamefont{E.}~\bibnamefont{Poisson}},
  \bibinfo{journal}{Phys. Rev.} \textbf{\bibinfo{volume}{D74}},
  \bibinfo{pages}{044009} (\bibinfo{year}{2006}), \eprint{gr-qc/0605077}.

\bibitem[{\citenamefont{Regge and {Wheeler}}(1957)}]{Regge:1957rw}
\bibinfo{author}{\bibfnamefont{T.}~\bibnamefont{Regge}} \bibnamefont{and}
  \bibinfo{author}{\bibfnamefont{J.~A.} \bibnamefont{{Wheeler}}},
  \bibinfo{journal}{Phys. Rev.} \textbf{\bibinfo{volume}{108}},
  \bibinfo{pages}{1063} (\bibinfo{year}{1957}).

\bibitem[{\citenamefont{Davis et~al.}(1972)\citenamefont{Davis, Ruffini, and
  Tiomno}}]{Davis:1972pa}
\bibinfo{author}{\bibfnamefont{M.}~\bibnamefont{Davis}},
  \bibinfo{author}{\bibfnamefont{R.}~\bibnamefont{Ruffini}}, \bibnamefont{and}
  \bibinfo{author}{\bibfnamefont{J.}~\bibnamefont{Tiomno}},
  \bibinfo{journal}{\prd} \textbf{\bibinfo{volume}{5}}, \bibinfo{pages}{2932}
  (\bibinfo{year}{1972}).

\bibitem[{\citenamefont{{Detweiler}}(1978)}]{Detweiler:1978aj}
\bibinfo{author}{\bibfnamefont{S.~L.} \bibnamefont{{Detweiler}}},
  \bibinfo{journal}{\apj} \textbf{\bibinfo{volume}{225}}, \bibinfo{pages}{687}
  (\bibinfo{year}{1978}).

\bibitem[{\citenamefont{{Detweiler} and {Szedenits}}(1979)}]{Detweiler:1979ds}
\bibinfo{author}{\bibfnamefont{S.~L.} \bibnamefont{{Detweiler}}}
  \bibnamefont{and}
  \bibinfo{author}{\bibfnamefont{E.}~\bibnamefont{{Szedenits}},
  \bibfnamefont{Jr.}}, \bibinfo{journal}{\apj} \textbf{\bibinfo{volume}{231}},
  \bibinfo{pages}{211} (\bibinfo{year}{1979}).

\bibitem[{\citenamefont{Cutler et~al.}(1994)\citenamefont{Cutler, Kennefick,
  and Poisson}}]{Cutler:1994pb}
\bibinfo{author}{\bibfnamefont{C.}~\bibnamefont{Cutler}},
  \bibinfo{author}{\bibfnamefont{D.}~\bibnamefont{Kennefick}},
  \bibnamefont{and} \bibinfo{author}{\bibfnamefont{E.}~\bibnamefont{Poisson}},
  \bibinfo{journal}{Phys. Rev.} \textbf{\bibinfo{volume}{D50}},
  \bibinfo{pages}{3816} (\bibinfo{year}{1994}).

\bibitem[{\citenamefont{Poisson}(1995)}]{Poisson:1995vs}
\bibinfo{author}{\bibfnamefont{E.}~\bibnamefont{Poisson}},
  \bibinfo{journal}{Phys. Rev.} \textbf{\bibinfo{volume}{D52}},
  \bibinfo{pages}{5719} (\bibinfo{year}{1995}), \eprint{gr-qc/9505030}.

\bibitem[{\citenamefont{Poisson}(1997)}]{Poisson:1997ad}
\bibinfo{author}{\bibfnamefont{E.}~\bibnamefont{Poisson}},
  \bibinfo{journal}{Phys. Rev.} \textbf{\bibinfo{volume}{D55}},
  \bibinfo{pages}{7980} (\bibinfo{year}{1997}).

\bibitem[{\citenamefont{Martel}(2004)}]{Martel:2003jj}
\bibinfo{author}{\bibfnamefont{K.}~\bibnamefont{Martel}},
  \bibinfo{journal}{Phys. Rev.} \textbf{\bibinfo{volume}{D69}},
  \bibinfo{pages}{044025} (\bibinfo{year}{2004}), \eprint{gr-qc/0311017}.

\bibitem[{\citenamefont{Barack and Lousto}(2005)}]{Barack:2005nr}
\bibinfo{author}{\bibfnamefont{L.}~\bibnamefont{Barack}} \bibnamefont{and}
  \bibinfo{author}{\bibfnamefont{C.~O.} \bibnamefont{Lousto}}
  (\bibinfo{year}{2005}), \eprint{gr-qc/0510019}.

\bibitem[{\citenamefont{Sopuerta and Laguna}(2006)}]{Sopuerta:2005gz}
\bibinfo{author}{\bibfnamefont{C.~F.} \bibnamefont{Sopuerta}} \bibnamefont{and}
  \bibinfo{author}{\bibfnamefont{P.}~\bibnamefont{Laguna}},
  \bibinfo{journal}{Phys. Rev.} \textbf{\bibinfo{volume}{D73}},
  \bibinfo{pages}{044028} (\bibinfo{year}{2006}), \eprint{gr-qc/0512028}.

\bibitem[{\citenamefont{Haas}(2007)}]{Haas:2007kz}
\bibinfo{author}{\bibfnamefont{R.}~\bibnamefont{Haas}}, \bibinfo{journal}{Phys.
  Rev.} \textbf{\bibinfo{volume}{D75}}, \bibinfo{pages}{124011}
  (\bibinfo{year}{2007}), \eprint{0704.0797}.

\bibitem[{\citenamefont{Vega and Detweiler}(2008)}]{Vega:2007mc}
\bibinfo{author}{\bibfnamefont{I.}~\bibnamefont{Vega}} \bibnamefont{and}
  \bibinfo{author}{\bibfnamefont{S.}~\bibnamefont{Detweiler}},
  \bibinfo{journal}{Phys. Rev.} \textbf{\bibinfo{volume}{D77}},
  \bibinfo{pages}{084008} (\bibinfo{year}{2008}), \eprint{0712.4405}.

\bibitem[{\citenamefont{Barack and Sago}(2007)}]{Barack:2007tm}
\bibinfo{author}{\bibfnamefont{L.}~\bibnamefont{Barack}} \bibnamefont{and}
  \bibinfo{author}{\bibfnamefont{N.}~\bibnamefont{Sago}},
  \bibinfo{journal}{Phys. Rev.} \textbf{\bibinfo{volume}{D75}},
  \bibinfo{pages}{064021} (\bibinfo{year}{2007}), \eprint{gr-qc/0701069}.

\bibitem[{\citenamefont{Burko and Khanna}(2007)}]{Burko:2006ua}
\bibinfo{author}{\bibfnamefont{L.~M.} \bibnamefont{Burko}} \bibnamefont{and}
  \bibinfo{author}{\bibfnamefont{G.}~\bibnamefont{Khanna}},
  \bibinfo{journal}{Europhys. Lett.} \textbf{\bibinfo{volume}{78}},
  \bibinfo{pages}{60005} (\bibinfo{year}{2007}), \eprint{gr-qc/0609002}.

\bibitem[{\citenamefont{Sundararajan et~al.}(2007)\citenamefont{Sundararajan,
  Khanna, and Hughes}}]{Sundararajan:2007jg}
\bibinfo{author}{\bibfnamefont{P.~A.} \bibnamefont{Sundararajan}},
  \bibinfo{author}{\bibfnamefont{G.}~\bibnamefont{Khanna}}, \bibnamefont{and}
  \bibinfo{author}{\bibfnamefont{S.~A.} \bibnamefont{Hughes}},
  \bibinfo{journal}{Phys. Rev.} \textbf{\bibinfo{volume}{D76}},
  \bibinfo{pages}{104005} (\bibinfo{year}{2007}), \eprint{gr-qc/0703028}.

\bibitem[{\citenamefont{Sundararajan et~al.}(2008)\citenamefont{Sundararajan,
  Khanna, Hughes, and Drasco}}]{Sundararajan:2008zm}
\bibinfo{author}{\bibfnamefont{P.~A.} \bibnamefont{Sundararajan}},
  \bibinfo{author}{\bibfnamefont{G.}~\bibnamefont{Khanna}},
  \bibinfo{author}{\bibfnamefont{S.~A.} \bibnamefont{Hughes}},
  \bibnamefont{and} \bibinfo{author}{\bibfnamefont{S.}~\bibnamefont{Drasco}},
  \bibinfo{journal}{Phys. Rev.} \textbf{\bibinfo{volume}{D78}},
  \bibinfo{pages}{024022} (\bibinfo{year}{2008}), \eprint{0803.0317}.

\bibitem[{\citenamefont{Sopuerta et~al.}(2005)\citenamefont{Sopuerta, Sun,
  Laguna, and Xu}}]{Sopuerta:2005rd}
\bibinfo{author}{\bibfnamefont{C.~F.} \bibnamefont{Sopuerta}},
  \bibinfo{author}{\bibfnamefont{P.}~\bibnamefont{Sun}},
  \bibinfo{author}{\bibfnamefont{P.}~\bibnamefont{Laguna}}, \bibnamefont{and}
  \bibinfo{author}{\bibfnamefont{J.}~\bibnamefont{Xu}} (\bibinfo{year}{2005}),
  \eprint{gr-qc/0507112}.

\bibitem[{\citenamefont{Barack and Golbourn}(2007)}]{Barack:2007jh}
\bibinfo{author}{\bibfnamefont{L.}~\bibnamefont{Barack}} \bibnamefont{and}
  \bibinfo{author}{\bibfnamefont{D.~A.} \bibnamefont{Golbourn}},
  \bibinfo{journal}{Phys. Rev.} \textbf{\bibinfo{volume}{D76}},
  \bibinfo{pages}{044020} (\bibinfo{year}{2007}), \eprint{0705.3620}.

\bibitem[{\citenamefont{Brandt and Bruegmann}(1997)}]{Brandt:1997tf}
\bibinfo{author}{\bibfnamefont{S.}~\bibnamefont{Brandt}} \bibnamefont{and}
  \bibinfo{author}{\bibfnamefont{B.}~\bibnamefont{Bruegmann}},
  \bibinfo{journal}{Phys. Rev. Lett.} \textbf{\bibinfo{volume}{78}},
  \bibinfo{pages}{3606} (\bibinfo{year}{1997}), \eprint{gr-qc/9703066}.

\bibitem[{\citenamefont{Campanelli et~al.}(2006)\citenamefont{Campanelli,
  Lousto, Marronetti, and Zlochower}}]{Campanelli:2005dd}
\bibinfo{author}{\bibfnamefont{M.}~\bibnamefont{Campanelli}},
  \bibinfo{author}{\bibfnamefont{C.~O.} \bibnamefont{Lousto}},
  \bibinfo{author}{\bibfnamefont{P.}~\bibnamefont{Marronetti}},
  \bibnamefont{and}
  \bibinfo{author}{\bibfnamefont{Y.}~\bibnamefont{Zlochower}},
  \bibinfo{journal}{Phys. Rev. Lett.} \textbf{\bibinfo{volume}{96}},
  \bibinfo{pages}{111101} (\bibinfo{year}{2006}), \eprint{gr-qc/0511048}.

\bibitem[{\citenamefont{Baker et~al.}(2006)\citenamefont{Baker, Centrella,
  Choi, Koppitz, and van Meter}}]{Baker:2005vv}
\bibinfo{author}{\bibfnamefont{J.~G.} \bibnamefont{Baker}},
  \bibinfo{author}{\bibfnamefont{J.}~\bibnamefont{Centrella}},
  \bibinfo{author}{\bibfnamefont{D.-I.} \bibnamefont{Choi}},
  \bibinfo{author}{\bibfnamefont{M.}~\bibnamefont{Koppitz}}, \bibnamefont{and}
  \bibinfo{author}{\bibfnamefont{J.}~\bibnamefont{van Meter}},
  \bibinfo{journal}{Phys. Rev. Lett.} \textbf{\bibinfo{volume}{96}},
  \bibinfo{pages}{111102} (\bibinfo{year}{2006}), \eprint{gr-qc/0511103}.

\bibitem[{\citenamefont{Barack et~al.}(2007)\citenamefont{Barack, Golbourn, and
  Sago}}]{Barack:2007we}
\bibinfo{author}{\bibfnamefont{L.}~\bibnamefont{Barack}},
  \bibinfo{author}{\bibfnamefont{D.~A.} \bibnamefont{Golbourn}},
  \bibnamefont{and} \bibinfo{author}{\bibfnamefont{N.}~\bibnamefont{Sago}},
  \bibinfo{journal}{Phys. Rev.} \textbf{\bibinfo{volume}{D76}},
  \bibinfo{pages}{124036} (\bibinfo{year}{2007}), \eprint{0709.4588}.

\bibitem[{\citenamefont{Lousto and Nakano}(2008)}]{Lousto:2008mb}
\bibinfo{author}{\bibfnamefont{C.~O.} \bibnamefont{Lousto}} \bibnamefont{and}
  \bibinfo{author}{\bibfnamefont{H.}~\bibnamefont{Nakano}},
  \bibinfo{journal}{Class. Quant. Grav.} \textbf{\bibinfo{volume}{25}},
  \bibinfo{pages}{145018} (\bibinfo{year}{2008}), \eprint{0802.4277}.

\bibitem[{\citenamefont{Canizares and Sopuerta}(2008)}]{Canizares:2008dp}
\bibinfo{author}{\bibfnamefont{P.}~\bibnamefont{Canizares}} \bibnamefont{and}
  \bibinfo{author}{\bibfnamefont{C.~F.} \bibnamefont{Sopuerta}}, in
  \emph{\bibinfo{booktitle}{Proceedings of the 7th International LISA
  Symposium}} (\bibinfo{year}{2008}), \eprint{0811.0294}.

\bibitem[{\citenamefont{Boyd}(2001)}]{Boyd}
\bibinfo{author}{\bibfnamefont{J.~P.} \bibnamefont{Boyd}},
  \emph{\bibinfo{title}{Chebyshev and Fourier Spectral Methods}}
  (\bibinfo{publisher}{Dover}, \bibinfo{address}{New York},
  \bibinfo{year}{2001}), \bibinfo{edition}{2nd} ed.

\bibitem[{\citenamefont{Grandclement and Novak}(2009)}]{Grandclement:2007sb}
\bibinfo{author}{\bibfnamefont{P.}~\bibnamefont{Grandclement}}
  \bibnamefont{and} \bibinfo{author}{\bibfnamefont{J.}~\bibnamefont{Novak}},
  \bibinfo{journal}{Living Rev. Relativity} \textbf{\bibinfo{volume}{12}},
  \bibinfo{pages}{1} (\bibinfo{year}{2009}), \eprint{0706.2286},
  \urlprefix\url{http://www.livingreviews.org/lrr-2009-1}.

\bibitem[{\citenamefont{Jung et~al.}(2007)\citenamefont{Jung, Khanna, and
  Nagle}}]{Jung:2007zf}
\bibinfo{author}{\bibfnamefont{J.-H.} \bibnamefont{Jung}},
  \bibinfo{author}{\bibfnamefont{G.}~\bibnamefont{Khanna}}, \bibnamefont{and}
  \bibinfo{author}{\bibfnamefont{I.}~\bibnamefont{Nagle}}
  (\bibinfo{year}{2007}), \eprint{0711.2545}.

\bibitem[{\citenamefont{Field et~al.}(2009)\citenamefont{Field, Hesthaven, and
  Lau}}]{Field:2009kk}
\bibinfo{author}{\bibfnamefont{S.~E.} \bibnamefont{Field}},
  \bibinfo{author}{\bibfnamefont{J.~S.} \bibnamefont{Hesthaven}},
  \bibnamefont{and} \bibinfo{author}{\bibfnamefont{S.~R.} \bibnamefont{Lau}}
  (\bibinfo{year}{2009}), \eprint{0902.1287}.

\bibitem[{\citenamefont{Quinn}(2000)}]{Quinn:2000wa}
\bibinfo{author}{\bibfnamefont{T.~C.} \bibnamefont{Quinn}},
  \bibinfo{journal}{Phys. Rev.} \textbf{\bibinfo{volume}{D62}},
  \bibinfo{pages}{064029} (\bibinfo{year}{2000}), \eprint{gr-qc/0005030}.

\bibitem[{\citenamefont{Barack and Ori}(2003)}]{Barack:2002bt}
\bibinfo{author}{\bibfnamefont{L.}~\bibnamefont{Barack}} \bibnamefont{and}
  \bibinfo{author}{\bibfnamefont{A.}~\bibnamefont{Ori}},
  \bibinfo{journal}{Phys. Rev.} \textbf{\bibinfo{volume}{D67}},
  \bibinfo{pages}{024029} (\bibinfo{year}{2003}), \eprint{gr-qc/0209072}.

\bibitem[{\citenamefont{Abramowitz and Stegun}(1972)}]{Abramowitz:1970as}
\bibinfo{author}{\bibfnamefont{M.}~\bibnamefont{Abramowitz}} \bibnamefont{and}
  \bibinfo{author}{\bibfnamefont{I.~A.} \bibnamefont{Stegun}},
  \emph{\bibinfo{title}{Handbook of Mathematical Functions with Formulas,
  Graphs, and Mathematical Tables}} (\bibinfo{publisher}{Dover},
  \bibinfo{address}{New York}, \bibinfo{year}{1972}).

\bibitem[{\citenamefont{Gustafsson et~al.}(1995)\citenamefont{Gustafsson,
  Kreiss, and Oliger}}]{Gustafsson:1995tb}
\bibinfo{author}{\bibfnamefont{B.}~\bibnamefont{Gustafsson}},
  \bibinfo{author}{\bibfnamefont{H.}~\bibnamefont{Kreiss}}, \bibnamefont{and}
  \bibinfo{author}{\bibfnamefont{J.}~\bibnamefont{Oliger}},
  \emph{\bibinfo{title}{Time dependent problems}} (\bibinfo{publisher}{John
  Wiley \& Sons}, \bibinfo{address}{New York}, \bibinfo{year}{1995}).

\bibitem[{\citenamefont{Frigo and Johnson}(2005)}]{fftw:2005}
\bibinfo{author}{\bibfnamefont{M.}~\bibnamefont{Frigo}} \bibnamefont{and}
  \bibinfo{author}{\bibfnamefont{S.~G.} \bibnamefont{Johnson}},
  \bibinfo{journal}{Proceedings of the IEEE} \textbf{\bibinfo{volume}{93}},
  \bibinfo{pages}{216} (\bibinfo{year}{2005}), \bibinfo{note}{special issue on
  "Program Generation, Optimization, and Platform Adaptation"}.

\bibitem[{\citenamefont{Butcher}(2008)}]{Butcher:2008jb}
\bibinfo{author}{\bibfnamefont{J.~C.} \bibnamefont{Butcher}},
  \emph{\bibinfo{title}{Numerical Methods for Ordinary Differential Equations}}
  (\bibinfo{publisher}{John Wiley \& Sons}, \bibinfo{address}{Chichester},
  \bibinfo{year}{2008}), \bibinfo{edition}{2nd} ed.

\bibitem[{\citenamefont{Press et~al.}(1992)\citenamefont{Press, Flannery,
  Teukolsky, and Vetterling}}]{Press:1992nr}
\bibinfo{author}{\bibfnamefont{W.~H.} \bibnamefont{Press}},
  \bibinfo{author}{\bibfnamefont{B.~P.} \bibnamefont{Flannery}},
  \bibinfo{author}{\bibfnamefont{S.~A.} \bibnamefont{Teukolsky}},
  \bibnamefont{and} \bibinfo{author}{\bibfnamefont{W.~T.}
  \bibnamefont{Vetterling}}, \emph{\bibinfo{title}{Numerical Recipes: The Art
  of Scientific Computing}} (\bibinfo{publisher}{Cambridge University Press},
  \bibinfo{address}{Cambridge}, \bibinfo{year}{1992}).

\bibitem[{\citenamefont{Hesthaven}(2000)}]{Hesthaven:2000jh}
\bibinfo{author}{\bibfnamefont{J.~S.} \bibnamefont{Hesthaven}},
  \bibinfo{journal}{Applied Numerical Mathematics}
  \textbf{\bibinfo{volume}{33}}, \bibinfo{pages}{23} (\bibinfo{year}{2000}).

\bibitem[{\citenamefont{Galassi et~al.}(2006)\citenamefont{Galassi, Davies,
  Theiler, Gough, Jungman, Booth, and Rossi}}]{Galasi:2006mg}
\bibinfo{author}{\bibfnamefont{M.}~\bibnamefont{Galassi}},
  \bibinfo{author}{\bibfnamefont{J.}~\bibnamefont{Davies}},
  \bibinfo{author}{\bibfnamefont{J.}~\bibnamefont{Theiler}},
  \bibinfo{author}{\bibfnamefont{B.}~\bibnamefont{Gough}},
  \bibinfo{author}{\bibfnamefont{G.}~\bibnamefont{Jungman}},
  \bibinfo{author}{\bibfnamefont{M.}~\bibnamefont{Booth}}, \bibnamefont{and}
  \bibinfo{author}{\bibfnamefont{F.}~\bibnamefont{Rossi}},
  \emph{\bibinfo{title}{GNU Scientific Library Reference Manual}}
  (\bibinfo{publisher}{Network Theory Ltd.}, \bibinfo{address}{Bristol},
  \bibinfo{year}{2006}), \bibinfo{edition}{2nd} ed.

\bibitem[{\citenamefont{Diaz-Rivera et~al.}(2004)\citenamefont{Diaz-Rivera,
  Messaritaki, Whiting, and Detweiler}}]{DiazRivera:2004ik}
\bibinfo{author}{\bibfnamefont{L.~M.} \bibnamefont{Diaz-Rivera}},
  \bibinfo{author}{\bibfnamefont{E.}~\bibnamefont{Messaritaki}},
  \bibinfo{author}{\bibfnamefont{B.~F.} \bibnamefont{Whiting}},
  \bibnamefont{and}
  \bibinfo{author}{\bibfnamefont{S.}~\bibnamefont{Detweiler}},
  \bibinfo{journal}{Phys. Rev.} \textbf{\bibinfo{volume}{D70}},
  \bibinfo{pages}{124018} (\bibinfo{year}{2004}), \eprint{gr-qc/0410011}.

\bibitem[{\citenamefont{Barack et~al.}(2008)\citenamefont{Barack, Ori, and
  Sago}}]{Barack:2008ms}
\bibinfo{author}{\bibfnamefont{L.}~\bibnamefont{Barack}},
  \bibinfo{author}{\bibfnamefont{A.}~\bibnamefont{Ori}}, \bibnamefont{and}
  \bibinfo{author}{\bibfnamefont{N.}~\bibnamefont{Sago}},
  \bibinfo{journal}{Phys. Rev.} \textbf{\bibinfo{volume}{D78}},
  \bibinfo{pages}{084021} (\bibinfo{year}{2008}), \eprint{0808.2315}.

\bibitem[{\citenamefont{Kosloff and Tal-Ezer}(1993)}]{Kosloff:1993kt}
\bibinfo{author}{\bibfnamefont{D.}~\bibnamefont{Kosloff}} \bibnamefont{and}
  \bibinfo{author}{\bibfnamefont{H.}~\bibnamefont{Tal-Ezer}},
  \bibinfo{journal}{Journal of Computational Physics}
  \textbf{\bibinfo{volume}{104}}, \bibinfo{pages}{457} (\bibinfo{year}{1993}).

\end{thebibliography}

\end{document}